\documentstyle[12pt,psfig]{article} \textwidth=17cm \textheight=22.5cm
\def\beq{\begin{equation}} \def\eeq{\end{equation}} \def\bea{\begin{eqnarray}}
\def\eea{\end{eqnarray}}
\def\bq{\begin{quote}} \def\eq{\end{quote}}

\def\gappeq{\mathrel{\rlap {\raise.5ex\hbox{$>$}} {\lower.5ex\hbox{$\sim$}}}}

\def\lappeq{\mathrel{\rlap{\raise.5ex\hbox{$<$}} {\lower.5ex\hbox{$\sim$}}}}

\begin{document} \pagestyle{empty} \begin{flushright} {CERN-TH/97-290}\\
{SNS/PH/1997-7}\\October 1997 \end{flushright} \vspace*{5mm}
\begin{center} {\bf ELECTROWEAK PRECISION TESTS: A CONCISE REVIEW\footnote{
This work is partially supported by the "Beyond the Standard Model" TMR
network under the EEC contract No. ERBFMRX-CT960090.}} \\
 \vspace*{1cm}  {\bf
G.Altarelli} \\ \vspace{0.2cm} Theoretical Physics Division, CERN
 \\ CH - 1211 Geneva 23 \\ and
\\Universit\`a di Roma Tre, Rome, Italy \\ \vspace{0.2cm} {\bf R. Barbieri}
 \\ \vspace{0.2cm}Scuola
Normale Superiore, Pisa, Italy\\ and \\INFN, Sezione di Pisa, Italy
\\\vspace{0.2cm} {\bf F. Caravaglios}\\
\vspace{0.2cm}
Theoretical Physics Division, CERN \\ CH - 1211 Geneva 23
\\ \vspace*{0.2cm}   {\bf Content} \\\end{center}
 \vspace*{0.cm} 1.Introduction\\2.Status of the Data \\3.Precision
Electroweak Data and the Standard Model\\4.A More General
 Analysis of Electroweak Data\\\indent 4.1
Basic Definitions and Results\\\indent 4.2 Experimental
 Determination of the Epsilon Variables\\
\indent 4.3
Comparing the Data with the Minimal Supersymmetric
 Standard Model\\5.Theoretical Limits on the Higgs
Mass\\6.Conclusion\\
\begin{center}\vspace*{-0.3cm} Submitted to Int. Journal of Modern Physics A
\\ 
\end{center}
\vspace*{-1cm}  \noindent  

\noindent

\begin{flushleft} {CERN-TH/97-290}\\
{SNS/PH/1997-7} \\ October 1997    \end{flushleft}
\vfill\eject 

\setcounter{page}{1} \pagestyle{plain}


\section{ Introduction}

In recent years new powerful tests of the Standard Model (SM) have been
 performed mainly at LEP but also
at SLC and at the Tevatron. The running of LEP1 was terminated in 1995 and
 close-to-final results of the data
analysis are now available \cite{tim},\cite{ew}. The experiments at the $Z_0$
 resonance have enormously improved
the accuracy in the electroweak neutral current sector. The top quark has been
 at last
found and  the errors on $m_Z$ and $\sin^2\theta_{eff}$ went down by two and
 one orders of magnitude,
respectively, since the start of LEP in 1989. 
The LEP2 programme is in progress.
 The validity
of the SM has been confirmed to a level that we can say was unexpected at the
 beginning. In the present data
there is no significant evidence for departures from the SM, no convincing
 hint of new physics (also
including the first  results from LEP2) \cite{dio}. The impressive success
 of the SM poses strong limitations on
the possible forms of  new physics. Favoured are models of the Higgs sector
 and of new physics that preserve the
SM structure  and only very delicately improve it, as is the case for
 fundamental Higgs(es) and
Supersymmetry. Disfavoured are models with a nearby strong non perturbative
 regime that  almost inevitably
would affect the radiative corrections, as for composite Higgs(es) or
 technicolor  and its variants\cite{30}-\cite{32}.

\section{ Status of the Data}

The relevant electro-weak data together with their SM values are presented
 in table 1 \cite{tim}-\cite{ew}.  The
SM predictions correspond to a fit of all the available data (including the
 directly measured values of $m_t$
and
$m_W$) in terms of $m_t$, $m_H$ and $\alpha_s(m_Z)$, described later in
 sect.3, table 4 (last column).

Other important derived quantities are, for example, $N_\nu$ the number of
 light neutrinos,
obtained from the invisible width: $N_\nu=2.993(11)$, which shows that only
 three fermion generations
exist with $m_\nu <45$ GeV, or the leptonic width $\Gamma_l$, averaged over
 e, $\mu$ and $\tau$:
$\Gamma_l= 83.91(10) MeV$, or the hadronic width: 
$\Gamma_h=1743.2(2.3)$ MeV.

For indicative purposes, in table 1 the "pulls" are also shown, defined as:
 pull = (data point- fit
value)/(error on data point).
At a glance we see that the agreement with the SM is quite good. The
 distribution of the
pulls is statistically normal. The presence of a few $\sim2\sigma$
 deviations is what is to be expected.
However it is maybe worthwhile to give a closer look at these small
 discrepancies.

Perhaps the most annoying feature of the data is the persistent difference
between the values of
$\sin^2\theta_{eff}$ measured at LEP and at SLC. The value of
 $\sin^2\theta_{eff}$ is obtained from a set of
combined asymmetries. From asymmetries one derives the ratio
 $x=g_V^l/g_A^l$ of the vector and axial vector
couplings of the $Z_0$, averaged over the charged leptons. In turn
 $\sin^2\theta_{eff}$ is
defined by $x=1-4\sin^2\theta_{eff}$. SLD obtains x from the single
 measurement of
$A_{LR}$, the left-right asymmetry, which requires longitudinally
 polarized beams. The distribution of the
present measurements of
$\sin^2\theta_{eff}$ is shown in fig. 1. The LEP average,
 $\sin^2\theta_{eff}=0.23199(28)$, differs by
$2.9\sigma$ from the SLD value
$\sin^2\theta_{eff}=0.23055(41)$. The most
precise individual measurement at LEP is from $A^{FB}_b$: the combined
 LEP error on this quantity is about the
same as the SLD error, but the two values are $3.1\sigma$'s away.
 One might attribute this to the fact that
the b measurement is more delicate and affected by a complicated
 systematics. In fact one notices from fig. 1
that the value  obtained at LEP from $A^{FB}_l$, the average for
 l=e, $\mu$ and $\tau$, is somewhat low (indeed
quite in agreement with the SLD value). However the statement that LEP
 and SLD agree on leptons while they only
disagree when the b quark is considered is not quite right. First, the value of
$A_e$, a quantity essentially identical to
$A_{LR}$, measured at LEP from the angular distribution of the $\tau$
 polarization, differs by
$1.8\sigma$ from the SLD value. Second, the low value of
$\sin^2\theta_{eff}$ found at LEP from
$A^{FB}_l$  turns out to be entirely due to the $\tau$ lepton channel which
 leads to a central value different
than that of e and
$\mu$ \cite{ew}. The e and $\mu$ asymmetries, which are experimentally simpler,
 are perfectly on top
of the SM fit. Suppose we take only e and $\mu$ asymmetries at LEP and
 disregard the b and $\tau$
measurements: the LEP average becomes $\sin^2\theta_{eff}=0.23184(55)$,
 which is still $1.9\sigma$ away
 from the SLD value.

In conclusion, it is difficult to find a simple explanation for the SLD-LEP
 discrepancy on
$\sin^2\theta_{eff}$. In view of this, the error on the nominal SLD-LEP
 average,
$\sin^2\theta_{eff}=0.23152(23)$, should perhaps be enlarged, for example,
 by a factor
$S=\sqrt{\chi^2/N_{df}}\sim1.4$, according to the recipe
adopted by the Particle Data Group in such cases.
Accordingly, in the following we will often use the
average
\beq \sin^2\theta_{eff}=0.23152\pm0.00032 \label{8} \eeq
Thus the LEP-SLC discrepancy results in an effective limitation of the
 experimental precision on
$\sin^2\theta_{eff}$. The data-taking by the SLD experiment is still in
 progress and also at LEP
seizable improvements on
$A_{\tau}$ and $A^{FB}_b$ are foreseen as soon as the corresponding
 analyses will be completed. We
hope to see the difference to decrease or to be understood.

From the above discussion one may wonder if there is evidence for
 something special in the $\tau$
channel, or equivalently if lepton universality is really supported by the
 data. Indeed this is the case: the
hint of a difference in $A^{FB}_\tau$ with respect to the corresponding
 e and
$\mu$ asymmetries is not confirmed by the measurements of  $A_\tau$ and
 $\Gamma_\tau$ which appear normal
\cite{tim},\cite{ew},\cite{li}. In principle the fact that an anomaly shows
 up in $A^{FB}_\tau$  and not in
$A_\tau$ and
$\Gamma_\tau$ is not unconceivable because the FB lepton asymmetries are
 very small and very precisely measured.
For example, the extraction of
$A^{FB}_\tau$ from the data on the angular distribution of $\tau$'s could
 be biased if the
imaginary part of the continuum was altered by some non universal new physics
 effect \cite{car}. But a more
trivial experimental problem is at the moment quite plausible.

A similar question can be asked for the b couplings. We have seen that the
 measured value of $A^{FB}_b$
is about $2\sigma$'s below the SM fit. At the same time $R_b$ which used to
 show a major discrepancy is
now only about $1.4\sigma$'s away from the SM fit (as a result of the more
 sophisticated second
generation experimental techniques).    It is often stated that there is a
 $-2.5\sigma$ deviation
on  the measured value of $A_b$ vs the SM
expectation \cite{tim},\cite{ew}. But in fact that  depends on how the data
 are combined. In our
opinion one should rather talk of a $-1.8\sigma$ effect. Let us discuss this
 point in detail.
\begin{table} Table 1: The electroweak data and the SM values obtained from 
a global fit.\\
\vglue.3cm
\begin{tabular}{|l|l|l|l|}
\hline Quantity&Data (August '97)       & Standard Model Fit & Pull\\
\hline
$m_Z$ (GeV)&91.1867(20) &91.1866 &~~0.0\\
$\Gamma_Z$ (GeV)        &2.4948(25) & 2.4966 & $-0.7 $\\
$\sigma_h$ (nb) &41.486(53)     &41.467 & ~~0.4\\
$R_h$   &20.775(27)     &20.756 & ~~0.7\\
$R_b$ &0.2170(9)       &0.2158 & ~~1.4\\
$R_c$&  0.1734(48)&     0.1723 & $-0.1$ \\
$A^l_{FB}$&  0.0171(10) &0.0162 & ~~0.9 \\
$A_\tau$ &      0.1411(64)      &0.1470 & $-0.9$ \\
$A_e$   &0.1399(73) &0.1470& $-1.0$\\
$A^b_{FB}$ &    0.0983(24) &0.1031 & $-2.0$ \\
$A^c_{FB}$&     0.0739(48)      &0.0736 & ~~0.0\\
$A_b$ (SLD direct)   & 0.900(50) &0.935 & $-0.7$\\
$A_c$ (SLD direct)  &  0.650(58) &0.668 & $-0.3$\\
$\sin^2\theta_{eff}({\rm\hbox{LEP-combined}})$ & 0.23199(28) & 0.23152& ~~1.7\\
$A_{LR}\rightarrow  \sin^2\theta_{eff}$& 0.23055(41) &  0.23152 & $-2.4$ \\
$m_W$ (GeV) (LEP2+p$\bar p$) & 80.43(8)       &80.375& ~~0.7\\
$1-\frac{m^2_W}{m^2_Z}$ ($\nu$N) &  0.2254(37) &0.2231 & $~~0.6$\\
$Q_W$ (Atomic PV in Cs) &  -72.11(93) &-73.20 & $~~1.2$\\
$m_t$ (GeV)     &175.6(5.5) &173.1 & ~~0.4\\
\hline
\end{tabular}
\end{table}
$A_b$ can be measured directly at SLC by taking advantage of the beam
 longitudinal polarization. At LEP
one measures
$A^{FB}_b$    = 3/4 $A_eA_b$. One can then derive $A_b$ by inserting a
 value for $A_e$. The question is what
to use for $A_e$: the LEP value obtained, using lepton
universality, from the measurements of $A^{FB}_l$, $A_\tau$, $A_e$:
 $A_e$ = 0.1461(33), or the
combination of LEP and SLD etc. The LEP
electroweak working group adopts for
$A_e$ the SLD+LEP average value which also includes $A_{LR}$ from SLD:
 $A_e$ = 0.1505(23). This procedure
leads to a $-2.5\sigma$ deviation. However, in this case,
the well known $\sim 2\sigma$ discrepancy of $A_{LR}$ with respect to $A_e$
 measured at LEP and also to the
SM fit, which is not related to the b couplings, further contributes to
 inflate the number of $\sigma$'s.
Since we are here concerned with the b couplings it is perhaps wiser to
 obtain $A_b$ from
LEP by using the SM value for
$A_e$ (that is the pull-zero value of table 1):
$A^{SM}_e$   = 0.1467(16). With the value of $A_b$ derived in this way
 from LEP we finally obtain \beq
 A_b = 0.895\pm0.022~~~~~(\rm{LEP+SLD, A_e=A^{SM}_e: -1.8}) \label{7}
\eeq In the SM $A_b$ is so close to 1
because the b quark is almost purely left-handed. $A_b$ only depends on
the ratio $r=(g_R/g_L)^2$ which in
the SM is small: $r\sim 0.033$. To adequately decrease $A_b$ from its SM
 value one must increase r by a factor
of about 1.6, which appears large for a new physics effect. Also such a
 large change in r must be compensated
by decreasing $g_L^2$ by a small but fine-tuned amount in order to
 counterbalance the corresponding
large positive shift in $R_b$. In view of this the most likely way out is
that $A^{FB}_b$ and
$A_b$ have been a bit underestimated at LEP and actually there is no anomaly
 in the b couplings. Then the LEP
value of $\sin^2\theta_{eff}$ would slightly move towards the SLD value, but,
 as explained above, by far not
enough to remove the SLD-LEP discrepancy (for example, if the LEP average for
 $\sin^2\theta_{eff}$ is
computed by moving the central value of $A^{FB}_b$ to the pull-zero value in
 table 1 with the same
figure for the error, one finds $\sin^2\theta_{eff}=0.23162(28)$, a value
 still $2.2\sigma$'s away from the 
SLD result).

\section{ Precision Electroweak Data and the Standard Model}

        For the analysis of electroweak data in the SM one starts from the
 input parameters: some of them,
$\alpha$, $G_F$ and $m_Z$, are very well measured, some other ones,
 $m_{f_{light}}$, $m_t$ and
$\alpha_s(m_Z)$  are only approximately determined while $m_H$ is largely
 unknown. With respect to
$m_t$ the situation has much improved since the CDF/D0 direct measurement
 of the top quark mass
\cite{gir}. From the input parameters one computes the radiative corrections
\cite{radcorr} to a
sufficient precision to match the experimental capabilities. Then one
compares the theoretical predictions
and the data for the numerous observables which have been measured, checks
 the consistency of the
theory and derives constraints on $m_t$, $\alpha_s(m_Z)$ and hopefully also
on $m_H$.

        Some comments on the least known of the input parameters are now in
 order. The only practically
relevant terms where precise values of the light quark masses,
 $m_{f_{light}}$, are needed are those
related to the hadronic contribution to the photon vacuum polarization
 diagrams, that determine
$\alpha(m_Z)$. This correction is of order 6$\%$, much larger than the
 accuracy of a few permil of
the precision tests. Fortunately, one can use the actual data to in
 principle solve the related
ambiguity. But the leftover uncertainty is still one of the main sources of
theoretical error. In recent years there has been a lot of activity on this
 subject and a number of
independent new estimates of $\alpha(m_Z)$  have appeared in the literature
 \cite{alfaQED},(see also
\cite{piet}). A consensus has been established and the value used at present is
\beq
\alpha(m_Z)^{-1}=128.90\pm0.09 \label{8a} \eeq

        As for the strong coupling $\alpha_s(m_Z)$ the world average central
value is by now quite stable. 
\begin{table}
\begin{center} Table 2: Measurements of $\alpha_s(m_Z)$. In
 parenthesis we indicate if the dominant source
of errors is theoretical or experimental. For theoretical ambiguities
 our personal figure of merit is given.\\
\vglue.3cm
\begin{tabular}{|l|ll|}
\hline Measurements & \multicolumn{2}{c|}{$\alpha_s(m_Z)$}\\
\hline
$R_{\tau}$ & 0.122 $\pm$ 0.006 & (Th)\\ Deep Inelastic Scattering & 0.116 $\pm$
0.005 & (Th)\\
$Y_{\rm decay}$ & 0.112 $\pm$ 0.010 & (Th)\\ Lattice QCD & 0.117 $\pm$ 0.007 &
(Th)\\
$Re^+e^-(\sqrt s < 62~{\rm GeV}$) & 0.124 $\pm$ 0.021 & (Exp)\\ Fragmentation
functions in $e^+e^-$ & 0.124 $\pm$ 0.012 & (Th)\\ Jets in $e^+e^-$ at
 and below
the $Z$ & 0.121 $\pm$ 0.008 & (Th)\\
$Z$ line shape (Assuming SM) & 0.120 $\pm$ 0.004 & (Exp)\\
\hline
\end{tabular}
\end{center}
\end{table}
The
error is going down because the dispersion among the different measurements
is much smaller in the most
recent set of data. The most important determinations of $\alpha_s(m_Z)$ are
 summarized in table 2 \cite{cat}. For
all entries, the main sources of error are the theoretical ambiguities which
are larger than the experimental
errors. The only exception is the measurement from the electroweak precision
 tests, but only if one assumes
that the SM electroweak sector is correct. Our personal views on the
 theoretical errors are reflected in
the table 2. The error on the final average is taken by all authors between
$\pm$0.003 and
$\pm$0.005 depending on how conservative one is. Thus in the following our
 reference value
will be \beq
\alpha_s(m_Z) = 0.119\pm0.004 \label{9} \eeq
        Finally a few words on the current status of the direct measurement
 of $m_t$. The present combined CDF/D0
result is \cite{gir}\beq
m_t = 175.6\pm 5.5~ {\rm GeV} \label{10} \eeq
The error is so small by now that one is approaching a level
where a more careful investigation of the effects of colour rearrangement on
 the determination of $m_t$ is
needed. One wants to determine the top quark mass, defined as the invariant
 mass of its decay products (i.e.
b+W+ gluons +
$\gamma$'s). However, due to the need of colour rearrangement, the top quark
 and its decay products cannot be
really isolated from the rest of the event. Some smearing of the mass
 distribution is induced by this colour
crosstalk which involves the decay products of the top, those of the antitop
 and also the fragments of the
incoming (anti)protons. A reliable quantitative computation of the
 smearing effect on the $m_t$
determination is difficult because of the importance of non perturbative
 effects. An induced error of
the order of one GeV on $m_t$ is reasonably expected. Thus further progress
on the $m_t$
determination demands tackling this problem in more depth.

        In order to appreciate the relative importance of the different
 sources of theoretical errors for
precision tests of the SM, we report in table 3  a comparison for the most
 relevant observables,
evaluated using refs. \cite{radcorr},\cite{radcorr2}.   What is important to
 stress is that the ambiguity from $m_t$,
once by far the largest one, is by now smaller than the error from $m_H$. We
also see from table 3 that the error
from
$\Delta\alpha(m_Z)$ is especially important for $\sin^2\theta_{eff}$  and, to
 a lesser extent, is also sizeable for
$\Gamma_Z$ and $\epsilon_3$, to be defined later on.
\begin{table} Table 3: Errors from different sources: $\Delta^{exp}_{now}$
   is
the present experimental error;
$\Delta\alpha^{-1}$ is the impact of $\Delta\alpha^{-1}=\pm0.09$;
  $\Delta_{th}$
is the estimated theoretical error from higher orders; $\Delta m_t$ is from
$\Delta m_t =\pm 6 $GeV;
$\Delta m_H$ is from $\Delta m_H$ = 60--1000 GeV; $\Delta \alpha_s$
 corresponds to
$\Delta \alpha_s=\pm0.003$. The epsilon parameters are defined in
sect. 4.1 \cite{abc}.
\begin{center}
\begin{tabular}{|l|l|l|l|l|l|l|}
\hline Parameter& $\Delta^{exp}_{now}$ & $\Delta \alpha^{-1}$ & $\Delta_{th}$ &
$\Delta m_t$ & $\Delta m_H$ & $\Delta \alpha_s$ \\
\hline
$\Gamma_Z$ (MeV) & $\pm$2.5 & $\pm$0.7 & $\pm$0.8 & $\pm$1.4 & $\pm$4.6 &
$\pm$1.7 \\
$\sigma_h$ (pb) & 53 & 1 & 4.3 & 3.3 & 4 & 17\\
$R_h \cdot 10^3$ & 27 & 4.3 & 5 & 2 & 13.5 & 20 \\
$\Gamma_l$ (keV) & 100 & 11 & 15 & 55 & 120 & 3.5\\
$A^l_{FB}\cdot 10^4$ & 10 & 4.2 & 1.3 & 3.3 & 13 & 0.18 \\
$\sin^2\theta\cdot 10^4$ & $\sim$3.2 & 2.3 & 0.8 & 1.9 & 7.5 & 0.1\\
$m_W$~(MeV) & 80 & 12 & 9 & 37 & 100& 2.2 \\
$R_b \cdot 10^4$ & 9 & 0.1 & 1 & 2.1 & 0.25 & 0\\
$\epsilon_1\cdot 10^3$ & 1.2 & & $\sim$0.1 & & & 0.2\\
$\epsilon_3\cdot 10^3$ & 1.4 & 0.5 & $\sim$0.1 & & & 0.12\\
$\epsilon_b\cdot 10^3$ & 2.1 & & $\sim$0.1 & & & 1\\
\hline
\end{tabular}
\end{center}
\end{table}

The most important recent advance in the theory of radiative corrections is
 the calculation of the
$o(g^4m^2_t/m^2_W)$ terms in $\sin^2\theta_{eff}$ and $m_W$ (not yet in
 $\delta\rho$) \cite{deg}. The result implies
a small but visible correction to the predicted values but expecially a
 seizable decrease of the ambiguity
from scheme dependence (a typical effect of truncation).

        We now discuss fitting the data in the SM. Similar studies based
 on older sets of data are found in
refs.\cite{fits}. As the mass of the top quark is finally rather precisely
known from CDF and D0 one must
distinguish two different types of fits. In one type one wants to answer the
 question: is $m_t$ from radiative
corrections in agreement with the direct measurement at the Tevatron? For
 answering this interesting but somewhat
limited question, one must clearly exclude the CDF/D0 measurement of $m_t$
 from the input set of data. Fitting all
other data in terms of
$m_t$,
$m_H$ and
$\alpha_s(m_Z)$ one finds the results shown in the third column of table 4
 \cite{ew}. Other similar fits
where also $m_W$ direct data are left out are shown.
\begin{table} Table 4: SM fits from different sets of data (with and without 
the direct measurements of $m_W$ and $m_t$).\\
\vglue.3cm
\begin{tabular}{|l|l|l|l|l|}
\hline Parameter & LEP(incl.$m_W$) &All but $m_W$, $m_t$ &All but  $m_t$ &
 All Data\\
\hline
$m_t$ (GeV) & 158$+14-11$ & 157$+10-9$ & 161$+10-8$ & $173.1\pm5.4$\\
$m_H$ (GeV) & 83$+168-49$ & 41$+64-21$ & 42$+75-23$ & 115$+116-66$\\
$log[m_H({\rm GeV})]$ & 1.92$+0.48-0.39$ &  1.62$+0.41-0.31$ & 
 1.63$+0.44-0.33$ &
 2.06$+0.30-0.37$\\
$\alpha_s(m_Z)$ & $0.121\pm0.003$ & $0.120\pm0.003$ & $0.120\pm0.003$ &
 $0.120\pm0.003$ \\
$\chi^2/dof$ & 8/9 & 14/12 & 16/14 & 17/15\\
\hline
\end{tabular}
\end{table}
The extracted value of $m_t$ is typically a bit too low. There is a strong
 correlation between $m_t$ and $m_H$.
The results on $\sin^2\theta_{eff}$ and $m_W$ \cite{kim} drive the fit to
 small values of
 $m_H$. This can be seen from
figs.2 and 3 (note that in fig. 2 the value of $\sin^2\theta_{eff}$ found by
 SLD would be too low to be shown on
the scale of the plot). Then, at small
$m_H$, the widths drive the fit to small $m_t$ (see fig. 4). In this context
 it is important to remark that
fixing
$m_H$ at 300 GeV, as was often done in the past, is by now completely
 obsolete, because it introduces too
strong a bias on the fitted value of
$m_t$. The change induced on the fitted value of $m_t$ when moving $m_H$
 from 300 to 65 or 1000 GeV is in
fact larger than the error on the direct measurement of $m_t$.

        In a more general type of fit, e.g. for determining the overall
 consistency of the SM or the best
present estimate for some quantity, say $m_W$, one should of course not
 ignore the existing direct
determinations of $m_t$ and $m_W$. Then, from all the available data,  by
 fitting
$m_t$, $m_H$ and $\alpha_s(m_Z)$ one finds the values shown in the last
 column of table 4.
This is the fit also referred to in table 1. The corresponding
fitted values of $\sin^2\theta_{eff}$ and $m_W$ are: \bea
\sin^2\theta_{eff} =0.23152\pm0.00022,\nonumber \\
                        m_W = 80.375\pm0.030 {\rm GeV} \label{10a} \eea The fitted
 value of $\sin^2\theta_{eff}$ is identical to the
LEP+SLD average and the caution on the error expressed in the previous
 section applies. The error of 30 MeV on
$m_W$  clearly sets up a goal for the direct measurement of $m_W$ at LEP2 and
 the Tevatron.

As a final comment we want to recall that the radiative corrections are
 functions of $log(m_H)$. It is truly
remarkable that the fitted value of $log(m_H)$ (the decimal logarithm)
 is found to fall right into
 the very narrow
allowed window around the value 2 specified by the lower limit from direct
 searches, $m_H>77$ GeV, and the
theoretical upper limit in the SM $m_H< 600-800$ GeV (see sect.6). The
 fulfilment of this very stringent consistency
check is a beautiful argument in favour of a fundamental Higgs
 (or one with a compositeness scale much above the
weak scale).

\section{A More General Analysis of Electroweak Data }

We now discuss an update of the epsilon analysis \cite{abc} which is a
 method to look at the data in
a more general context than the SM. The starting point is to isolate from
 the data that part which is due to the
purely weak radiative corrections. In fact the epsilon variables are
 defined in such a way that they are zero
in the approximation when only effects from the SM at tree level plus
 pure QED and pure QCD corrections
are taken into account. This very simple version of improved Born
 approximation is a good first
approximation  according to the data and is independent of $m_t$ and $m_H$. In
fact the whole $m_t$ and $m_H$ dependence arises from weak loop corrections
 and therefore is only contained
in the epsilon variables. Thus the epsilons are extracted from the data
 without need of specifying
$m_t$ and $m_H$. But their predicted value in the SM or in any extension of
 it depend on $m_t$ and $m_H$.
This is to be compared with the competitor method based on the S, T, U
 variables \cite{pes},\cite{abar}. The latter
cannot be obtained from the data without specifying
$m_t$ and $m_H$ because they are defined as deviations from the complete SM
 prediction for specified $m_t$ and
$m_H$. Of course there are very many variables that vanish if pure weak loop
 corrections are neglected, at
least one for each relevant observable. Thus for a useful definition we
 choose a set of
representative observables that are used to parametrize those hot spots of
the radiative corrections where
new physics effects are most likely to show up. These sensitive weak
correction terms include vacuum
polarization diagrams which being potentially quadratically divergent are
 likely to contain all possible non
decoupling effects (like the quadratic top quark mass dependence in the SM).
 There are three independent
vacuum polarization contributions. In the same spirit, one must add the
 $Z\rightarrow b \bar b$ vertex which also
includes a large top mass dependence. Thus altogether we consider four
 defining observables: one asymmetry, for
example
$A_{FB}^l$, (as representative of the set of measurements that lead to the
 determination of
$\sin^2\theta_{eff}$), one width (the leptonic width
$\Gamma_l$ is particularly suitable because it is practically independent of
 $\alpha_s$), $m_W$ and $R_b$.
Here lepton universality has been taken for granted, because the data show
 that it is
verified within the present accuracy. The
four variables,
$\epsilon_1$, $\epsilon_2$, $\epsilon_3$ and $\epsilon_b$ are defined
 in ref.$\cite{abc}$ in one to one
correspondence with the set of observables  $A^{FB}_l$, $\Gamma_l$,
$m_W$, and $R_b$. The definition is so chosen that the quadratic top mass
 dependence is only
present  in
$\epsilon_1$ and
$\epsilon_b$, while the
$m_t$ dependence of
$\epsilon_2$ and $\epsilon_3$ is logarithmic. The definition of $\epsilon_1$
 and $\epsilon_3$ is specified
in terms of $A^{FB}_l$ and $\Gamma_l$ only. Then adding $m_W$ or $R_b$ one
 obtains $\epsilon_2$ or
$\epsilon_b$. We now specify the relevant definitions in detail.

\subsection{Basic Definitions and Results}

We start from the basic observables $m_W/m_Z$, $\Gamma_l$ and  $A^{FB}_l$ and
 $\Gamma_b$. From these four
quantities one can isolate the corresponding dynamically significant
 corrections $\Delta r_W$, $\Delta \rho$,
$\Delta k$ and $\epsilon_b$, which  contain the small effects one is trying to
 disentangle and are defined in
the following. First we introduce $\Delta r_W$ as obtained from $m_W/m_Z$ by
the relation:
\beq
(1-\frac{m_W^2}{m_Z^2}) \frac{m_W^2}{m_Z^2}~=~\frac{\pi \alpha(m_Z)}{\sqrt{2}
G_F m_Z^2 (1-\Delta r_W)}
\label{1n}
\eeq
Here $\alpha(m_Z)~=~\alpha /(1-\Delta \alpha)$ is fixed to the central value
 1/128.90 so that the effect of
the running of $\alpha$ due to known physics is extracted from $1-\Delta r =
 (1- \Delta \alpha)(1- \Delta
r_W)$. In fact, the error on $1/\alpha(m_Z)$, as given in eq.(\ref{8a}) would
then affect $\Delta r_W$.
In order to define $\Delta
\rho$ and
$\Delta k$ we
consider the effective vector and axial-vector couplings $g_V$ and $g_A$ of
 the on-shell Z to charged leptons,
given by the formulae:
\bea
\Gamma_l~=~\frac{G_F m^3_Z}{6\pi \sqrt{2}}(g^2_V+g_A^2)
 (1+\frac{3 \alpha}{4 \pi}), \nonumber \\
A_l^{FB}(\sqrt{s}=m_Z)~=~\frac{3g^2_Vg^2_A}{(g^2_V+g_A^2)^2}~=~\frac{3x^2}
{(1+x^2)^2}. \label{2n}
\eea
Note that $\Gamma_l$ stands for the inclusive partial width
 $\Gamma(Z\rightarrow l\bar l + \rm{photons})$. We
stress the following points. First, we have extracted from $(g^2_V+g_A^2)$
 the factor $(1 + 3\alpha /4 \pi )$
which is induced in $\Gamma_l$ from final state radiation. Second, by the
 asymmetry at the peak in
eq.(\ref{2n}) we mean the quantity which is commonly referred to by the LEP
 experiments (denoted as $A^0_{FB}$
in ref.\cite{ew}), which is corrected for all QED effects, including initial
 and final state radiation and
also for the effect of the imaginary part of the $\gamma$ vacuum polarization
  diagram. In terms of $g_A$ and
$x= g_V /g_A$, the quantities $\Delta \rho$ and
$\Delta k$
are given by:
\bea
g_A~=~-\frac{\sqrt{\rho}}{2}~\sim~-\frac{1}{2}(1+\frac{\Delta \rho}{2}),
 \nonumber \\
x~=~\frac{g_V}{g_A}~=~1-4\sin^2\theta_{eff}~=~1-4(1+\Delta k) s_0^2.\label{3n}
\eea
Here $s_0^2$ is $\sin^2\theta_{eff}$
before non pure-QED corrections, given by:
\beq
s_0^2 c_0^2~=~ \frac{\pi \alpha(m_Z)}{\sqrt{2} G_F m_Z^2} \label{4n}
\eeq
with  $c_0^2~=~1-s_0^2$ ($s_0^2 = 0.231095$ for $m_Z~=~91.188$ GeV).

We now define $\epsilon_b$ from $\Gamma_b$, the inclusive partial width for
 $Z\rightarrow b \bar b$ according to
the relation
\beq
\Gamma_b~=~\frac{G_F m^3_Z}{6\pi \sqrt{2}}\beta (\frac{3-\beta^2}{2}
 g^2_{bV}~+~\beta^2 g^2_{bA}) N_C R_{QCD}
(1+\frac{\alpha}{12\pi}) \label{5n}
\eeq
where $N_C=3$ is the number of colours, $\beta=\sqrt{1-4m_b^2/m^2_Z}$,
 with $m_b=4.7~$
GeV, $R_{QCD}$ is the QCD correction factor given by
\beq
R_{QCD}~=~ 1~+~1.2a~-~1.1a^2~-~13a^3~;~~~a~=~\frac{\alpha_s(m_Z)}{\pi}
 \label{6n}
\eeq
and $g_{bV}$ and $g_{bA}$ are specified as follows
\bea
                        g_{bA}~=~-\frac{1}{2}(1+\frac{\Delta \rho}{2})
(1+\epsilon_b), \nonumber\\
                        \frac{g_{bV}}{g_{bA}}~=~\frac{1-4/3\sin^2\theta_{eff}+
\epsilon_b}{1+\epsilon_b}.\label{7n}
\eea
This is clearly not the most general deviation from the SM in the
 $Z\rightarrow b \bar b$ vertex but $\epsilon_b$ is
closely related to the quantity  $-Re(\delta_{b-vertex})$ defined in the
 last of refs.\cite{zbb} where the large
$m_t$ corrections are located.

As is well known, in the SM the quantities $\Delta r_W$, $\Delta \rho$,
 $\Delta k$  and $\epsilon_b$, for
sufficiently  large $m_t$, are all dominated by  quadratic terms in $m_t$ of
 order $G_Fm^2_t$.   As new physics
can  more easily be disentangled if not masked by large conventional $m_t$
 effects, it is convenient to keep
$\Delta \rho$ and $\epsilon_b$ while trading $\Delta r_W$
and
$\Delta k$ for two quantities with no contributions of order $G_Fm^2_t$. We
 thus introduce the
following linear combinations:
\bea
\epsilon_1&=&\Delta \rho, \nonumber \\
\epsilon_2&=&c^2_0 \Delta \rho~+~\frac{s^2_0 \Delta r_W}{c^2_0-s^2_0}~-~2s^2_0
 \Delta k, \nonumber\\
\epsilon_3&=&c^2_0 \Delta \rho~+~(c^2_0-s^2_0) \Delta k. \label{8n}
\eea
The quantities $\epsilon_2$ and $\epsilon_3$ no longer contain terms of order
 $G_Fm^2_t$ but only logarithmic
terms in $m_t$. The leading terms for large Higgs mass, which are logarithmic,
 are contained in
$\epsilon_1$ and $\epsilon_3$. In the Standard Model one has the following
"large"
asymptotic contributions:
\bea
\epsilon_1&=&\frac{3G_F m_t^2}{8 \pi^2 \sqrt{2}}~-~\frac{3G_F m_W^2}{4 \pi^2
 \sqrt{2}} \tan^2{\theta_W}
\ln{\frac{m_H}{m_Z}}~+....,\nonumber \\
\epsilon_2&=&-\frac{G_F m_W^2}{2 \pi^2 \sqrt{2}}\ln{\frac{m_t}{m_Z}}~+....,
\nonumber \\
\epsilon_3&=&\frac{G_F m_W^2}
{12 \pi^2 \sqrt{2}}\ln{\frac{m_H}{m_Z}}~-~\frac{G_F m_W^2}{6 \pi^2
\sqrt{2}}\ln{\frac{m_t}{m_Z}}....,\nonumber \\
\epsilon_b&=&-\frac{G_F m_t^2}{4 \pi^2 \sqrt{2}}~+.... \label{9n}
\eea

The relations between the basic observables and the epsilons can be
 linearised, leading to the
approximate formulae
\bea
\frac{m_W^2}{m_Z^2}~=~\frac{m_W^2}{m_Z^2}\vert_B (1+ 1.43\epsilon_1 -
 1.00\epsilon_2 - 0.86\epsilon_3),
\nonumber \\
\Gamma_l~=~\Gamma_l\vert_B (1+ 1.20\epsilon_1 - 0.26\epsilon_3),
\nonumber \\
A_l^{FB}~=~A_l^{FB} \vert_B (1+ 34.72\epsilon_1 - 45.15\epsilon_3),
\nonumber \\
\Gamma_b~=~\Gamma_b\vert_B (1+ 1.42\epsilon_1 - 0.54\epsilon_3 +
 2.29\epsilon_b).  \label{10n}
\eea
The  Born approximations, as defined above, depend on $\alpha_s(m_Z)$
and also on $\alpha(m_Z)$. Defining
\beq
\delta \alpha_s~=~\frac{\alpha_s(m_Z)-0.119}{\pi};~~~\delta
\alpha~=~\frac{\alpha(m_Z)-\frac{1}{128.90}}{\alpha},~~~~ \label{11n}
\eeq
we have
\bea
\frac{m_W^2}{m_Z^2}\vert_B~=~0.768905(1-0.40\delta \alpha), \nonumber \\
\Gamma_l\vert_B~=~83.563(1-0.19\delta \alpha) \rm{MeV}, \nonumber \\
A_l^{FB} \vert_B~=~0.01696(1-34\delta \alpha), \nonumber \\
\Gamma_b\vert_B~=~379.8(1+1.0\delta \alpha_s-0.42\delta \alpha). \label{12n}
\eea
Note that the dependence on $\delta \alpha_s$ for $\Gamma_b\vert_B$, shown in
 eq.(\ref{12n}), is not simply the
one loop result for $m_b=0$ but a combined effective shift which takes into
 account both finite mass effects
and the contribution of the known higher order terms.

\begin{table} Table 5: Values of the epsilons in the SM as functions of $m_t$
 and
$m_H$ as obtained from recent versions\cite{radcorr2} of ZFITTER  and
 TOPAZ0 (also including the new results of ref.\cite{deg}).
These values (in
$10^{-3}$ units) are obtained for
$\alpha_s(m_Z)$ = 0.119,
$\alpha(m_Z)$ = 1/128.90, but the theoretical predictions are essentially
independent of
$\alpha_s(m_Z)$ and $\alpha(m_Z)$
 \cite{abc}.
\begin{center}
\begin{tabular}{|c|l|l|l|l|l|l|l|l|l|c|}
\hline
$m_t$ & \multicolumn{3}{|c|}{$\epsilon_1$}&\multicolumn{3}{|c|}{$\epsilon_2$}
&\multicolumn{3}{|c|}{$\epsilon_3$}&$\epsilon_b$\\ (GeV)& \multicolumn{3}{|c|}
{$m_H$ (GeV) =} &  \multicolumn{3}{|c|} {$m_H$ (GeV) =} & \multicolumn{3}{|c|}
{$m_H$ (GeV) =} & All {$m_H$}\\ & 70 & 300 & 1000 & 70 & 300 & 1000 & 70 & 300
 &
1000 &\\
\hline 150      &3.55&  2.86    & 1.72 &        $-$6.85 &       $-$6.46 &
   $-$5.95 &        4.98    & 6.22 &        6.81 &
$-$4.50 \\ 160 &        4.37 &  3.66 &  2.50 &  $-$7.12 &       $-$6.72 &
    $-$6.20 &       4.96 &   6.18 &
6.75 &  $-$5.31
\\
 170 &  5.26 &  4.52 &  3.32 &  $-$7.43 &        $-$7.01 &        $-$6.49 &
     4.94 &  6.14 &  6.69 &
$-$6.17\\
 180 &  6.19 &   5.42 &  4.18 &   $-$7.77 &       $-$7.35 &       $-$6.82 &
     4.91 &  6.09 &  6.61 &
$-$7.08\\
 190 &  7.18 &   6.35 &  5.09 &  $-$8.15 &       $-$7.75 &       $-$7.20 &
    4.89 &  6.03 &   6.52 &
$-$8.03\\
 200 &  8.22 &  7.34 &  6.04 &   $-$8.59 &       $-$8.18 &       $-$7.63 &
     4.87 &  5.97 &  6.43 &
$-$9.01\\
\hline
\end{tabular}
\end{center}
\end{table}

The important property of the epsilons is that, in the Standard Model, for all
 observables at the Z pole, the
whole dependence on $m_t$ (and $m_H$) arising from one-loop diagrams only
 enters through the epsilons. The same
is actually true, at the relevant level of precision, for all higher order
 $m_t$-dependent corrections.
Actually, the only residual $m_t$ dependence of the various observables not
 included in the epsilons is in the
terms of order $\alpha_s^2(m_Z)$ in the pure QCD correction factors to the
 hadronic widths \cite{kk}. But this
one is quantitatively irrelevant, especially in view of the errors connected
to the uncertainty on the value of
$\alpha_s(m_Z)$. The theoretical values of the epsilons in the SM from state
 of the art radiative corrections
\cite{radcorr, radcorr2}, also including the recent development of
 ref.\cite{deg}, are given in table 5. It is
important to remark that the theoretical values of the epsilons in
the SM, as given in table 2, are not
affected, at the percent level or so, by reasonable variations of
 $\alpha_s(m_Z)$ and/or $\alpha(m_Z)$ around
their central values. By our definitions, in fact,  no terms of
 order $\alpha_s^n(m_Z)$ or $\alpha \ln{m_Z/m}$
contribute to the epsilons.  In terms of the epsilons, the following
 expressions hold, within the SM, for the
various precision observables
\bea
\Gamma_T~=~\Gamma_{T0}(1+1.35\epsilon_1-0.46\epsilon_3+0.35\epsilon_b),
 \nonumber\\
R~=~R_0(1+0.28\epsilon_1-0.36\epsilon_3+0.50\epsilon_b), \nonumber\\
\sigma_h~=~\sigma_{h0}(1-0.03\epsilon_1+0.04\epsilon_3-0.20\epsilon_b),
 \nonumber\\
x~=~x_0(1+17.6\epsilon_1-22.9\epsilon_3), \nonumber\\
R_b~=~R_{b0}(1-0.06\epsilon_1+0.07\epsilon_3+1.79\epsilon_b). \label{13n}
\eea
where x=$g_V/g_A$ as obtained from $A_l^{FB}$ . The quantities in
 eqs.(\ref{10n},\ref{13n}) are clearly not
independent and the redundant information is reported for convenience.
 By comparison with the codes of
ref.\cite{radcorr2} (we also added the complete results of ref.\cite{deg})
 we obtain
\bea
\Gamma_{T0}~=~2489.46(1+0.73\delta \alpha_s-0.35\delta \alpha)~MeV,\nonumber \\
R_0~=~20.8228(1+1.05\delta \alpha_s-0.28\delta \alpha),\nonumber \\
\sigma_{h0}~=~41.420(1-0.41\delta \alpha_s+0.03\delta \alpha)~nb,\nonumber \\
x_0~=~0.075619-1.32\delta \alpha,\nonumber \\
R_{b0}~=~0.2182355.\label{14n}
\eea
Note that  the quantities in eqs.(\ref{14n}) should not be confused, at
 least in principle, with the
corresponding Born approximations, due to small "non universal" electroweak
corrections. In practice, at the
relevant level of approximation, the difference between the two corresponding
quantities is in any case
significantly smaller than the present experimental error.

In principle, any four observables could have been picked up as defining
 variables.
In practice we choose those that have a more clear physical significance and
 are more effective in the
determination of the epsilons. In fact,  since $\Gamma_b$ is actually
 measured by $R_b$ (which is nearly
insensitive to $\alpha_s$), it is preferable to use directly $R_b$  itself as
 defining variable, as we shall do
hereafter. In practice, since the value in eq.(\ref{14n}) is practically
 indistinguishable from the Born
approximation of $R_b$, this determines no change in any of the equations
 given above but simply requires the
corresponding replacement among the defining relations of the epsilons.

\subsection{Experimental Determination of the Epsilon Variables}

The values of the epsilons as obtained, following the
specifications in the previous sect.4.1, from the defining variables $m_W$,
 $\Gamma_l$, $A^{FB}_l$ and $R_b$
are shown in the first column of table 6.
\begin{table} Table 6: Experimental values of the epsilons in the SM from
different sets of data.
These values (in
$10^{-3}$ units) are obtained for
$\alpha_s(m_Z) = 0.119\pm0.003$,
$\alpha(m_Z) = 1/128.90\pm0.09$, the corresponding uncertainties being
included in the quoted errors.
\begin{tabular}{|l|l|l|l|l|}
\hline $\epsilon~~~10^3$  &Only def. quantities &All asymmetries &All High
Energy & All Data\\
\hline
$\epsilon_1~10^3$ &$4.0\pm1.2$ &$4.3\pm1.2$ &$4.1\pm1.2$  &$3.9\pm1.2$ \\
$\epsilon_2~10^3$ &$-8.3\pm2.3$ &$-9.1\pm2.2$ &$-9.3\pm2.2$  &$-9.4\pm2.2$ \\
$\epsilon_3~10^3$ &$2.9\pm1.9$ &$4.3\pm1.4$ &$4.1\pm1.4$  &$3.9\pm1.4$ \\
$\epsilon_b~10^3$ &$-3.2\pm2.3$ &$-3.3\pm2.3$ &$-3.9\pm2.1$  &$-3.9\pm2.1$  \\
\hline
\end{tabular}
\end{table}
To proceed further and include other measured observables in the analysis
 we need to make some
dynamical assumptions. The minimum amount  of model dependence is introduced
 by including other purely
leptonic quantities at the Z pole such as $A_{\tau}$, $A_e$ (measured  from
 the angular
dependence of the $\tau$ polarization) and $A_{LR}$ (measured by SLD). For
 this step, one is simply
assuming that the different leptonic asymmetries are equivalent measurements
 of $\sin^2\theta_{eff}$ (for an
example of a peculiar model where this is not true, see ref.\cite{carLR}) .
 We add, as usual, the measure of
$A^{FB}_b$ because this observable is dominantly sensitive to the leptonic
vertex. We then use the combined value
of $\sin^2\theta_{eff}$ obtained from the whole set of asymmetries measured
 at LEP and SLC with the error
increased according to eq.(\ref{8}) and the related discussion. At this stage
 the best values of the epsilons
are shown in the second column of table 6. In figs. 5-8  we report the
 1$\sigma$ ellipses in the indicated
$\epsilon_i$-$\epsilon_j$ planes that correspond to this set of
input data. In fig. 9, for example, we also give a graphical
representation in the $\epsilon_3$-$\epsilon_b$ plane, of the uncertainties
 due to $\alpha(m_Z)$ and
$\alpha_s(m_Z)$.

        All observables measured on the Z peak at LEP can be included in the
 analysis provided that we assume
that all deviations from the SM are only contained in vacuum polarization
diagrams (without demanding
a truncation of the $q^2$ dependence of the corresponding functions) and/or
 the $Z\rightarrow b\bar
b$  vertex. From a global fit of the data on $m_W$,  $\Gamma_T$,  $R_h$,
$\sigma_h$,  $R_b$ and
$\sin^2\theta_{eff}$ (for LEP data, we have taken the correlation matrix
for $\Gamma_T$,  $R_h$ and
$\sigma_h$ given by the LEP experiments \cite{ew}, while we have considered
 the additional information
on $R_b$ and $\sin^2\theta_{eff}$  as independent) we obtain the values shown
 in the third column of table
6. The comparison of theory and experiment at this stage is also shown in
 figs. 5-8. More detailed
information is shown in figs. 10-11, which both refer to the level when also
 hadronic data are taken
into account. But in fig. 10 we compare the results obtained if
$\sin^2\theta_{eff}$ is extracted in turn from
different asymmetries among those listed in fig. 1. The ellipse marked
 "average" is the same as the one
labeled "All high en." in fig. 6 and corresponds to the value of
 $\sin^2\theta_{eff}$ which is shown on the
figure (and in eq.(\ref{8})). We confirm that the value from $A_{LR}$ is
 far away from the SM given the
experimental value of $m_t$ and the bounds on $m_H$ and would correspond
 to very small values of
$\epsilon_3$ and of $\epsilon_1$. We see also that while the $\tau$ FB
 asymmetry is also on the low side,
the combined e and $\mu$ FB asymmetry are right on top of the average.
 Finally the b FB asymmetry is on the
high side. An analogous plot is presented in fig. 11. In this case the defining
 width $\Gamma_l$ is replaced
in turn with either the total or the hadronic or the invisible width. The
important conclusion that one
obtains is that the widths are indeed well consistent among them even with
 respect to this new criterium
of leading to the same epsilons.

         To include in our analysis lower energy observables as well, a
 stronger hypothesis needs to be
made:  vacuum polarization diagrams are allowed to vary from the SM  only
 in their constant and first
derivative terms in a $q^2$ expansion \cite{pes}-\cite{abar}. In such a case,
 one can, for example, add to the
analysis the ratio
$R_\nu$ of neutral to charged current processes in deep inelastic neutrino
 scattering on nuclei
\cite{33}, the "weak charge" $Q_W$  measured in atomic parity violation
experiments on Cs \cite{34}
and the measurement of $g_V/g_A$ from $\nu_\mu e$ scattering \cite{35}.
 In this way one obtains  the
global fit given in the fourth column of table 6 and shown in figs. 5-8.
For completeness, we also report the corresponding values of 
$\Delta r_W$ and $\Delta k$ (defined in eqs. (\ref{1n},\ref{3n})):
 $ 10^3 \times \Delta r_W=-27.0\pm4.3$, $10^3 \times \Delta k=16.7\pm 18.4$.   
 With the
progress of LEP the low energy data, while important as a check that no
 deviations from the expected
$q^2$ dependence arise, play a lesser role in the global fit. Note that
the present ambiguity on the
value of $\delta\alpha^{-1}(m_Z) =\pm0.09$ \cite{alfaQED} corresponds to
 an uncertainty on $\epsilon_3$ (the
other epsilons are not much affected) given by $\Delta\epsilon_3~10^3 =\pm0.6$
 \cite{abc}. Thus the
theoretical error is still confortably less than the experimental error. In
 fig. 12 we present a summary of the
experimental values of the epsilons as compared to the SM predictions as
 functions of $m_t$ and $m_H$, which
shows agreement within $1\sigma$. However the central values of $\epsilon_1$,
 $\epsilon_2$ and $\epsilon_3$ are all somewhat 
low, while the central value of $\epsilon_b$ is shifted upward with respect
to the SM as a
consequence of the  still imperfect matching of $R_b$.

A number of
interesting features are clearly visible from figs.5-12. First, the good
 agreement with the SM and the
evidence for weak corrections, measured by the distance of the data from the
improved Born approximation point
(based on tree level SM plus pure QED or QCD corrections). There is by now a
 solid evidence for departures from
the improved Born approximation where all the epsilons vanish. In other words
 a clear evidence for the pure
weak radiative corrections has been obtained and LEP/SLC are now measuring the
 various components of these
radiative corrections. For example, some authors \cite{39} have studied the
sensitivity of the data to a
particularly interesting subset of the weak radiative corrections, i.e. the
 purely bosonic part. These terms
arise from virtual exchange of gauge bosons and Higgses. The result is that
 indeed the measurements are
sufficiently precise to require the presence of these contributions in order
 to fit the data. Second, the
general results of the SM fits are reobtained from a different perspective.
 We see the preference for light
Higgs manifested by the tendency for
$\epsilon_3$ to be rather on the low side. Since $\epsilon_3$ is practically
 independent of $m_t$, its low value
demands $m_H$ small. If the Higgs is light then the preferred value of
$m_t$ is somewhat lower than the Tevatron result (which in the epsilon
 analysis is not included among the input
data). This is because also the value of $\epsilon_1\equiv \delta \rho$,
 which is determined by the widths, in
particular by the leptonic width, is somewhat low. In fact
$\epsilon_1$ increases with $m_t$ and, at fixed $m_t$, decreases with $m_H$,
 so that for small $m_H$ the low
central value of $\epsilon_1$ pushes $m_t$ down. Note that also the central
 value of $\epsilon_2$ is on
the low side, because the experimental value of $m_W$ is a little bit too
 large. Finally, we see that adding the
hadronic quantities or the low energy observables hardly makes a difference
 in the
$\epsilon_i$-$\epsilon_j$ plots with respect to the case with only the
 leptonic variables being included (the
ellipse denoted by "All Asymm."). But, take for example  the
$\epsilon_1$-$\epsilon_3$ plot: while the leptonic ellipse contains the same
information as one could obtain from a
$\sin^2\theta_{eff}$ vs $\Gamma_l$ plot, the content of the other two ellipses
 is much larger because it
shows that the hadronic as well as the low energy quantities match the
 leptonic variables without need of any new
physics. Note that the experimental values of $\epsilon_1$ and
$\epsilon_3$ when the hadronic quantities are included also depend on the
 input value of $\alpha_s$ given
in eq.(\ref{9}).

\subsection{Comparing the Data with the Minimal Supersymmetric Standard Model}

The MSSM \cite{43} is a completely specified,
consistent and computable theory. There are too many parameters to attempt a
 direct fit of the data to
the most general framework. So we consider two significant limiting cases:
 the "heavy" and the
"light" MSSM.

        The "heavy" limit corresponds to all sparticles being sufficiently
 massive, still within the limits
of a natural explanation of the weak scale of mass. In this limit a very
 important result holds
\cite{58}: for what concerns the precision electroweak tests, the MSSM
 predictions tend to reproduce
the results of the SM with a light Higgs, say $m_H\sim$ 100 GeV. So if the
masses of SUSY partners are pushed
at sufficiently large values the same quality of fit as for the SM is
 guaranteed. Note that for $m_t=175.6$ GeV
and $m_H\sim70$ GeV the values of the four epsilons computed in the SM lead
to a fit of the corresponding
experimental values with
$\chi^2\sim4$, which is reasonable for $d.o.f=4$. This value corresponds to
 the fact that the central values of
$\epsilon_1$,$\epsilon_2$, $\epsilon_3$ and -$\epsilon_b$ are all below the
SM value by about $1\sigma$, as can
be seen from fig. 12.

        In the "light" MSSM option some of the superpartners have a
 relatively small mass, close to their
experimental lower bounds. In this case the pattern of radiative corrections
may sizeably deviate from
that of the SM \cite{pok}. The potentially largest effects occur in vacuum
 polarisation amplitudes and/or the
$Z\rightarrow b\bar b$  vertex. In particular we recall the following
 contributions :

        i) a threshold effect in the Z wave function renormalisation \cite{58}
 mostly due to the vector
coupling of charginos and (off-diagonal) neutralinos to the Z itself.
 Defining the vacuum polarisation
functions by $\Pi_{\mu\nu}(q^2)=-ig_{\mu\nu}[A(0)+q^2 F(q^2)]+q_\mu q_\nu$
 terms, this is a positive
contribution to $\epsilon_5=m^2_Z  F'_{ZZ} (m^2_Z)$, the prime denoting a
 derivative with respect to
$q^2$ (i.e. a contribution to a higher derivative term not included in the
naive epsilon formalism, but compatible with the scheme described
in Sect. 4.1). The
$\epsilon_5$ correction shifts $\epsilon_1$, $\epsilon_2$ and $\epsilon_3$ by
-$\epsilon_5$,
-$c^2\epsilon_5$ and  -$c^2\epsilon_5$ respectively, where
 $c^2=\cos^2{\theta_W}$, so that all of them
are reduced by a comparable amount. Correspondingly all the Z widths are
 reduced without affecting the
asymmetries. This effect falls down particularly fast when the lightest
 chargino mass increases from a
value close to $m_Z$/2. Now that we know,  from the LEP2 runs, that the
 chargino mass is
probably not smaller than $m_Z$ its possible impact is drastically reduced.

        ii) a positive contribution to $\epsilon_1$ from the virtual exchange
 of split multiplets of SUSY partners,
for example of the scalar top and bottom  superpartners \cite{59}, analogous
 to the contribution of the
top-bottom left-handed quark doublet. From the experimental value of $m_t$ not
much space is left for this possibility, and the experimental value of
 $\epsilon_1$ is an important constraint
on the spectrum. This is especially true now that the rather large lower
limits on the chargino mass reduce the
size of a possible compensation from $\epsilon_5$. For example, if the stop
is light then it must be mainly a
right-handed stop. Also large values of $\tan\beta$ are disfavoured because
 they tend to enhance the splittings
among SUSY partner multiplets. In general it is simpler to decrease the
 predicted values of $\epsilon_2$ and
$\epsilon_3$ by taking advantage of $\epsilon_5$ than to decrease
 $\epsilon_1$, because the negative shift
from $\epsilon_5$ is most often counterbalanced by the increase from the
 effect of split SUSY multiplets.

        iii) a negative contribution to $\epsilon_b$ due to the virtual
 exchange of a charged Higgs
\cite{60}. If one defines, as customary, $\tan{\beta}=v_2/v_1$
($v_1$ and $v_2$ being the vacuum
expectation values of the Higgs doublets giving masses to the down and up
 quarks, respectively), then,
for negligible bottom Yukawa coupling or $\tan{\beta}<< m_t/m_b$, this
 contribution is proportional to
$m^2_t$ /$\tan^2{\beta}$.

        iv) a positive contribution to $\epsilon_b$ due to virtual
 chargino--stop exchange \cite{61} which
in this case is proportional to $m^2_t$ /$\sin^2{\beta}$ and prefers small
 $\tan\beta$. This effect
again requires the chargino and the  stop to be light in order to be sizeable.

With the recent limits set by LEP2 on the masses of SUSY partners the above
 effects are small enough that other
contributions from vertex diagrams could be comparable. Thus in the following
 we will only consider the
experimental values of the epsilons obtained at the level denoted by "All
 Asymmetries" which only assumes lepton
universality.

We have analysed the problem of what configurations of masses in the "light"
 MSSM are favoured or disfavoured
by the present data (updating ref.{\cite{63}). We find that no lower limits
 on the masses of SUSY partners are
obtained which are better than the direct limits. One exception is the case
 of stop and sbottom  masses, which
are severely constrained by the $\epsilon_1$ value and also, at small
 $\tan\beta$, by the increase at LEP2 of the
direct limit on the Higgs mass. Charged Higgs masses are also 
 constrained. Since the central
values of
$\epsilon_1$,$\epsilon_2$ and
$\epsilon_3$ are all below the SM it is convenient to make $\epsilon_5$ as
 large as possible. For this purpose
light gaugino and slepton masses are favoured. We find that for
$m_{\chi^+_1}\sim 90-120$ GeV the effect is
still sizeable. Also favoured are small values of $\tan\beta$ that allow to
 put slepton masses relatively low,
say, in the range 100-500~GeV, without making the split in the isospin
 doublets too large for $\epsilon_1$.
Charged Higgses must be heavy because they contribute to $\epsilon_b$ with
the wrong sign.  A light right-handed
stop could help on $R_b$ for a Higgsino-like chargino. But one needs small
 mixing (the right-handed stop must be
close to the mass eigenstate) and beware of the Higgs mass constraint at
 small $\tan\beta$ (a Higgs mass above
$\sim 80$ GeV, the range of LEP2 for susy Higgses at $\sqrt{s}=183$ GeV, 
starts being a strong
constraint at small $\tan\beta$). So we
prefer in the following to keep the stop mass large. The limits on
 $b\rightarrow s\gamma$ also prefer heavy charged Higgs
and stop \cite{zwi}.

The scatter plots obtained in the planes $\epsilon_1$-$\epsilon_3$ and
 $\epsilon_2$-$\epsilon_3$ for
$-200<\mu<200$ GeV, $0<M<250$ GeV, $\tan\beta=1.5-2.5$,
 $m_{\tilde l}=100-500$ GeV and $m_{\tilde q}= 1$ TeV
are shown in figs. 13 and 14, together with the SM prediction for
 $m_t=175.6$ GeV
and $m_h\sim70$ GeV. We see that in most cases the $\chi^2$ is not improved.
 If we restrict to the small area,
marked with a small star in fig. 13, 
where both  $\epsilon_1$ and $\epsilon_3$ are
 improved we can check that also
$\epsilon_2$ is improved, as is seen from fig. 14 where the same values of the
 parameters have been employed in the region marked with the star.
This region where the $\chi^2$ is decreased by slightly more than one unity is
 included in the hypervolume $\mu
=133-147$ GeV, $M=212-250$ GeV, ($m_{\chi^+_1}=90-105$ GeV,
 $m_{\chi^0_1}=58-72$ GeV,
$m_{\chi^0_2}=129-147$ GeV) with $\tan\beta\sim1.5$, $m_{\tilde q}\sim 1$ TeV
 and
$m_{\tilde l}=100$ GeV. In this configuration $\epsilon_b$ is unchanged. We
 see that the advantage with respect
to the SM is at most of the order of 1 in $\chi^2$.

\section{Theoretical Limits on the Higgs Mass}

The SM works with remarkable accuracy. But the experimental foundation of
the SM is not completed if the electroweak symmetry breaking mechanism is not
 experimentally established.
Experiments must decide what is true: the SM Higgs or Higgs plus SUSY or new
strong forces and Higgs
compositeness.

The theoretical limits on the Higgs mass play an important role in the
 planning of the experimental strategy.
The large experimental value of $m_t$ has important implications on
$m_H$ both in the minimal SM \cite{zziii}$-$\cite{bbiiii} and in its minimal
supersymmetric
extension\cite{cciiii}$,$\cite{ddiiii}.

        It is well known\cite{zziii}$-$\cite{bbiiii} that in the SM with only
 one Higgs doublet a lower limit on
$m_H$ can be derived from the requirement of vacuum stability. The limit is a
function of $m_t$ and of the
energy scale $\Lambda$ where the model breaks down and new physics appears.
 Similarly an upper bound on $m_H$
(with mild dependence on $m_t$) is obtained \cite{eeiiii} from the requirement
 that up to the scale $\Lambda$ no
Landau pole appears. If one demands vacuum stability up to a very large scale,
of the order of $M_{GUT}$ or
$M_{Pl}$ then the resulting bound on
$m_H$ in the SM with only one Higgs doublet is given by \cite{aaiiii}:
\begin{equation} m_H({\rm GeV}) > 138 + 2.1 \left[ m_t({\rm GeV}) -
 175.6 \right] -
 3.0~\frac{\alpha_s(m_Z) - 0.119}{0.004}~.
\label{25}
\end{equation}
In fact one can show that the discovery of a Higgs particle at
LEP2, or $m_H\lappeq 100$ GeV, would imply that the SM breaks down at a scale
$\Lambda$ of the order of a few TeV. Of course, the limit is only valid in the
 SM
with one doublet of Higgses. It is enough to add a second doublet to avoid the
 lower limit. The upper limit on
the Higgs mass in the SM is important for assessing the chances of success of
 the LHC as an accelerator designed
to solve the Higgs problem. The upper limit \cite{eeiiii} has been recently
reevaluated \cite{hr}. For $m_t\sim
175$ GeV one finds
$m_H\lappeq 180$ GeV for $\Lambda\sim M_{GUT}-M_{Pl}$ and $m_H\lappeq
 0.5-0.8$ TeV for $\Lambda\sim
1$ TeV.

A particularly
important example of a theory where the bound is violated is the MSSM, which
we now discuss.
As is well known \cite{43}, in the MSSM there are two Higgs doublets, which
 implies three neutral physical
Higgs particles and a pair of charged Higgses. The lightest neutral Higgs,
 called $h$, should be lighter than
$m_Z$ at tree-level approximation. However, radiative corrections
\cite{ffiiii} increase the $h$ mass by a term
proportional to $m^4_t$ and logarithmically dependent on the stop mass.
 Once the radiative corrections are
taken into account the $h$ mass still remains rather small: for
 $m_t$ = 174~GeV one finds the limit (for all
values of tg $\beta)~m_h < 130$~GeV
\cite{ddiiii}. Actually there are reasons to expect that $m_h$ is well
 below the bound. In fact, if $h_t$ is
large at the GUT scale, which is suggested by the large observed value
 ot $m_t$ and by a natural onsetting of
the electroweak symmetry breaking induced by $m_t$, then at low energy a
 fixed point is reached in the
evolution of $m_t$. The fixed point corresponds to $m_t \sim 195
 \sin\beta$~GeV (a good approximate relation
for $\tan\beta = v_{up}/v_{down} < 10$). If the fixed point situation is
realized, then $m_h$ is considerably
below the bound, $m_h\lappeq 100~GeV$ \cite{ddiiii}.

In conclusion, for $m_t \sim 175$~GeV, we have seen that, on the one hand, if
 a Higgs is found at LEP the SM
cannot be valid up to $M_{Pl}$. On the other hand, if a Higgs is found at LEP,
 then the MSSM has good chances,
because this model would be excluded for $m_h > 130$~GeV.

\section{Conclusion}

The experiments performed in recent years mainly at LEP but also at SLAC
and at the Tevatron have allowed to test the SM of the electroweak
interactions
with unprecedented precision. A number of observables measured at the per mille
level can be successfully fitted in terms of the most relevant parameters
of the
SM, $m_t, \alpha_S(m_Z)$ and $m_H$. The presence of a few $\sim2\sigma$
 deviations is what is to be expected on statistical grounds.
Furthermore, a closer look at such deviations 
does not give any hint of a significant pattern. An annoying
feature of the data is the persistent difference between the values of
$\sin^2\theta_{eff}$ measured at LEP and at SLC. There are reasons to think,
however, that
this difference will be understood by the further
data-taking at SLAC and by the completion of the LEP analyses of the $\tau$
and $b$ asymmetries.

A way to appreciate the significance of the attained precision is to notice,
as shown in table 3, that the uncertainties on most of the observables
are dominated, in the SM, by the variation of the Higgs mass from $60$ to
$1000$ GeV
and that such effect is often larger than the experimental
error. Remarkably, the fitted value of $\log(m_H)$,
which gets fixed in this way, falls right on top of the
range specified by the experimental lower limit and the theoretical upper
bound.
We interpret this as indirect evidence for a weakly interacting Higgs,
or Higgs plus SUSY, against new strong forces or an Higgs composite at
a light scale. This is a clear and easier conclusion than setting an
upper bound on the Higgs mass, within the SM, relevant to its search in
different machines, LEP, Tevatron or LHC. In this respect, it is only
unfortunate
that  $\sin^2\theta_{eff}$, which is among the most sensitive observables,
suffers of the problem pointed out before. In turn, this emphasizes the
importance of the
direct measurement of $m_W$ at LEP2 and the Tevatron, again
because of its sensitivity to
the Higgs mass.

The analysis of the data
in terms of the epsilon parameters shows the significance of
the precision tests in a more general context than the SM itself.
It would be most useful in the case that deviations were put in evidence.
This not being the case, this analysis serves mostly to illustrate
the evidence reached for quantum electroweak effects.
In the space of the $\epsilon$ parameters the region indicated by
 the data clearly excludes the Born point, i.e. the origin, by many
 $\sigma$'s. 
Since these parameters
are successfully fitted in the SM, any extension, or alternative of it
must be delicate enough not to undo this agreement. This is the case for
the MSSM, as clearly illustrated in fig.s 13, 14. For what concerns the
precision electroweak tests, the MSSM predictions reproduce the SM with
a light Higgs as soon as the sparticle masses are sufficiently heavy,
but still  compatible with the naturalness bound. A marginal reduction
of the $\chi^2$ is possible in a small region of the MSSM parameter space,
to which we cannot attach, however, at the moment any particular significance.

FIGURE CAPTIONS

1. The collected measurements of $\sin^2\theta_{eff}$. The resulting value for
 the $\chi^2$ is given by
$\chi^2/d.o.f=1.87$. As a consequence the error on the average is enlarged in
 the text by a factor $\sqrt{1.87}$
with respect to the formal average shown here.

2. SM prediction for  $\sin^2\theta_{eff}$ as function of $m_H$ for
 $m_t=175.6\pm5.5$ as computed from
updated radiative corrections \cite{deg}. The theoretical error bands
 from neglect of higher order terms, estimated
from scheme and scale dependence, are shown.The combined LEP+SLD experimental
 value is indicatively plotted for
$m_H\sim100$ GeV (the SLD value would be very low, out of the plot scale).
Small values of $m_H$ are preferred.

3. SM prediction for  $m_W$ as function of $m_H$ for $m_t=175.6\pm5.5$ as
 computed from
updated radiative corrections \cite{deg}. The theoretical error bands from
 neglect of higher order terms, estimated
from scheme and scale dependence, are shown. The combined LEP+hadron
 colliders experimental value is indicatively
plotted for $m_H\sim100$ GeV.Small values of $m_H$ are preferred.

4. SM prediction for the leptonic width as function of $m_t$. For small Higgs
 mass a value of $m_t$ slightly
smaller than the CDF/D0 experimental value is indicated.

5. Data vs theory in the $\epsilon_2$-$\epsilon_1$ plane. The origin point
 corresponds to the "Born"
approximation obtained from the SM at tree level plus pure QED and pure QCD
corrections. The predictions of the
full SM (also including the improvements of ref.\cite{deg}) are shown for
 $m_H$ = 70, 300 and 1000 GeV and
$m_t=175.6\pm5.5$ GeV (a segment for each $m_H$ with the arrow showing the
direction of
$m_t$ increasing from
$-1\sigma$ to $+1\sigma$). The three
$1-\sigma$ ellipses ($38\%$ probability contours) are obtained from a) "All
Asymm.": $\Gamma_l$, $m_W$ and
$\sin^2\theta_{eff}$ as obtained from the combined asymmetries (the value and
 error used are shown); b) "All High
En.": the same as in a) plus all the hadronic variables at the Z; c) "All
 Data": the same as in b) plus the low
energy data.

6. Data vs theory in the $\epsilon_3$-$\epsilon_1$ plane (notations as in
 fig. 5).

7. Data vs theory in the $\epsilon_2$-$\epsilon_3$ plane (notations as in
fig. 5).

8. Data vs theory in the $\epsilon_b$-$\epsilon_1$ plane (notations as in
fig. 5).

9. Data vs theory in the $\epsilon_3$-$\epsilon_b$ plane (notations as in
fig. 5, except that both ellipses
refer to the case b)) The inner $1-\sigma$ ellipse is without the errors
induced by the uncertainties on
$\alpha(m_Z)$ and $\alpha_s(m_Z)$.

10. Data vs theory in the $\epsilon_3$-$\epsilon_1$ plane (notations as in
fig. 5). The ellipse indicated with "Average"
corresponds to the case "All high en" of fig. 6 and is obtained from the
average value of $sin^2\theta_{eff}$ displayed
on the figure. The other ellipses are obtained by replacing the average
 $sin^2\theta_{eff}$ with the values obtained in
turn from each individual asymmetry as shown by the labels.

11. Data vs theory in the $\epsilon_3$-$\epsilon_1$ plane (notations as in
fig. 5). The different ellipses are obtained
from $m_W$, the average value of $sin^2\theta_{eff}$ displayed
on the figure, $R_b$ plus one width among $\Gamma_l$, $\Gamma_{inv}$, the
total width $\Gamma_Z$ and $\Gamma_h$.

12. The bands (labeled by the $\epsilon$ index) are the predicted values of
the epsilons in the SM as functions of
$m_t$ for
$m_H~=~70-1000$ GeV (the $m_H$ value corresponding to one edge of the band is
 indicated). The CDF/D0 experimental
1-$\sigma$ range of $m_t$ is shown. The experimental results for the epsilons
 from all data are displayed (from the last
column of table 6). The position of the data on the $m_t$ axis has been
 arbitrarily chosen and has no particular
meaning.

13.Scatter plot obtained in the plane $\epsilon_1$-$\epsilon_3$ for
$-200<\mu<200$ GeV, $0<M<250$ GeV, $\tan\beta=1.5-2.5$,
 $m_{\tilde l}=100-500$ GeV and $m_{\tilde q}= 1~TeV$,
together with the SM prediction for $m_t=175.6$ GeV and $m_H\sim70$ GeV.
The separation $\mu>0$ or $\mu<0$ is
clearly visible. The small star indicates the region with minimum $\chi^2$.

14.Scatter plot obtained in the plane $\epsilon_2$-$\epsilon_3$ for
$-200<\mu<200$ GeV, $0<M<250$ GeV, $\tan\beta=1.5-2.5$,
 $m_{\tilde l}=100-500$ GeV and $m_{\tilde q}= 1$ TeV,
together with the SM prediction for $m_t=175.6$ GeV and $m_H\sim70$ GeV.
The small star corresponds to the same values of the 
parameters  of  the region (marked with a star)  in fig. 13. 
For these values of parameters the fit of
 $\epsilon_1$,$\epsilon_2$ and
$\epsilon_3$ improves while $\epsilon_b$ is unchanged.

\pagestyle{empty}
\begin{figure}[t]
\psfig{figure=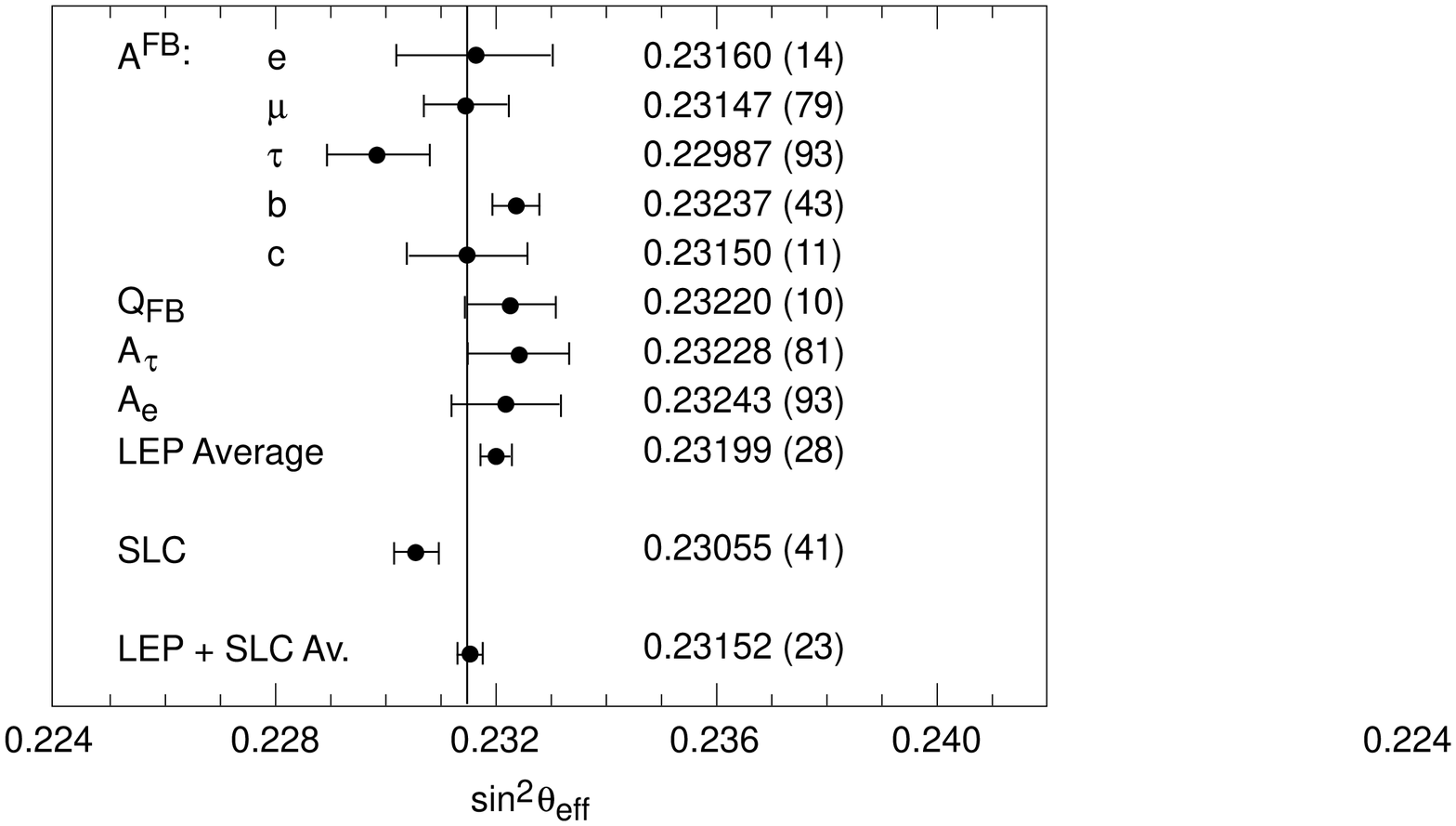}
\center{Fig. 1}
\end{figure}
\begin{figure}[t]
\psfig{figure=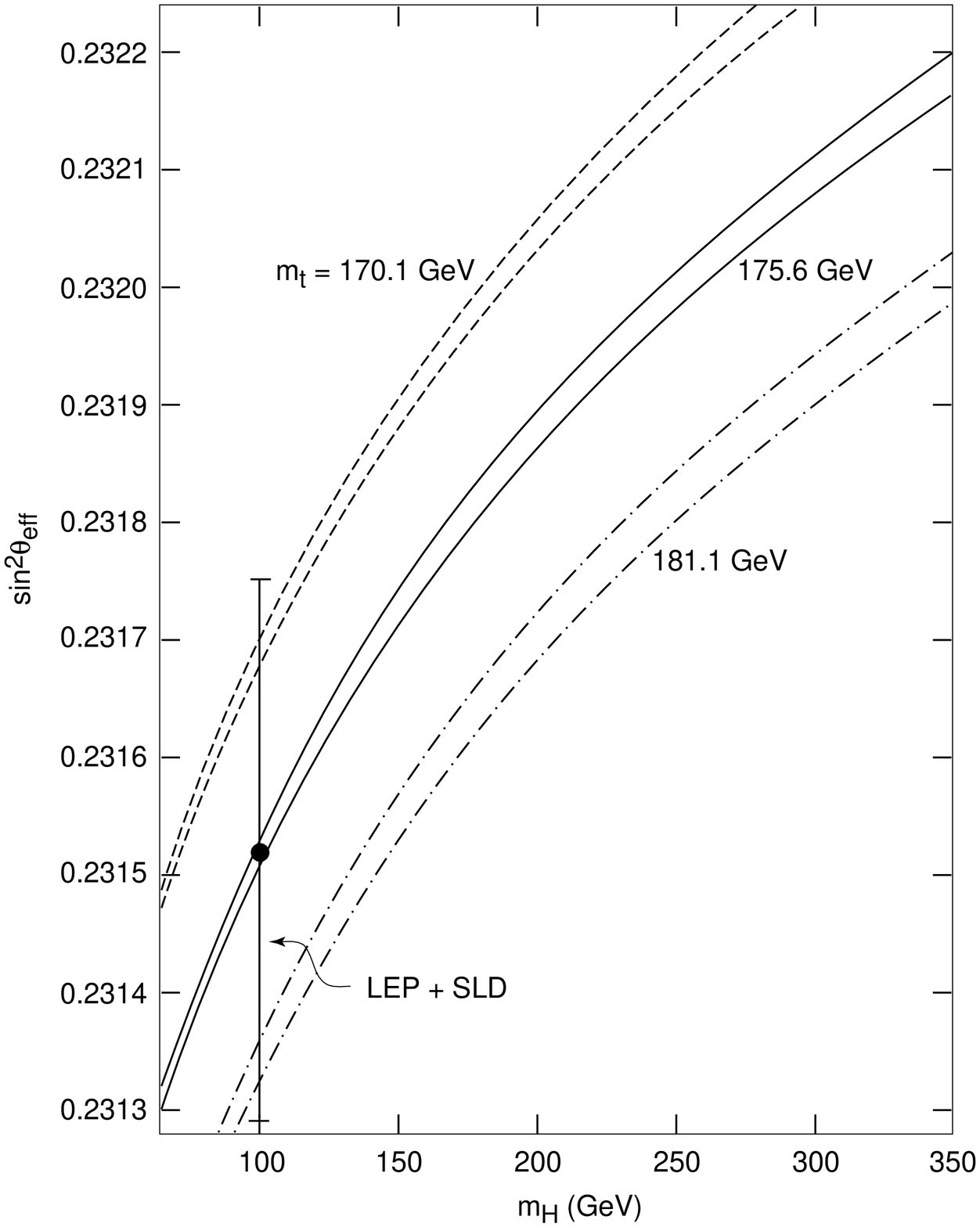,width=18cm}
\center{Fig. 2}
\end{figure}
\begin{figure}[t]
\psfig{figure=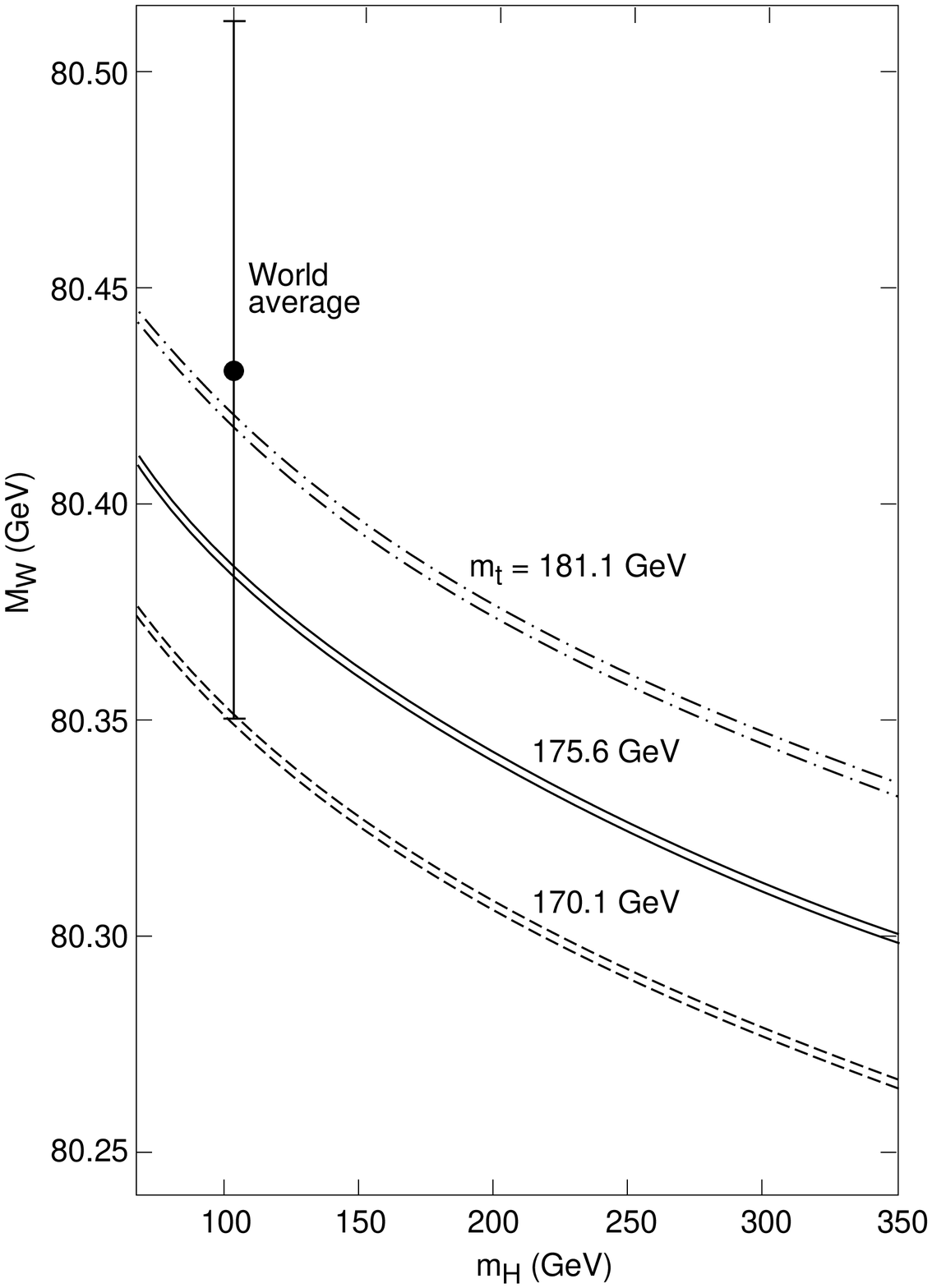}
\center{Fig. 3}
\end{figure}
\begin{figure}[t]
\psfig{figure=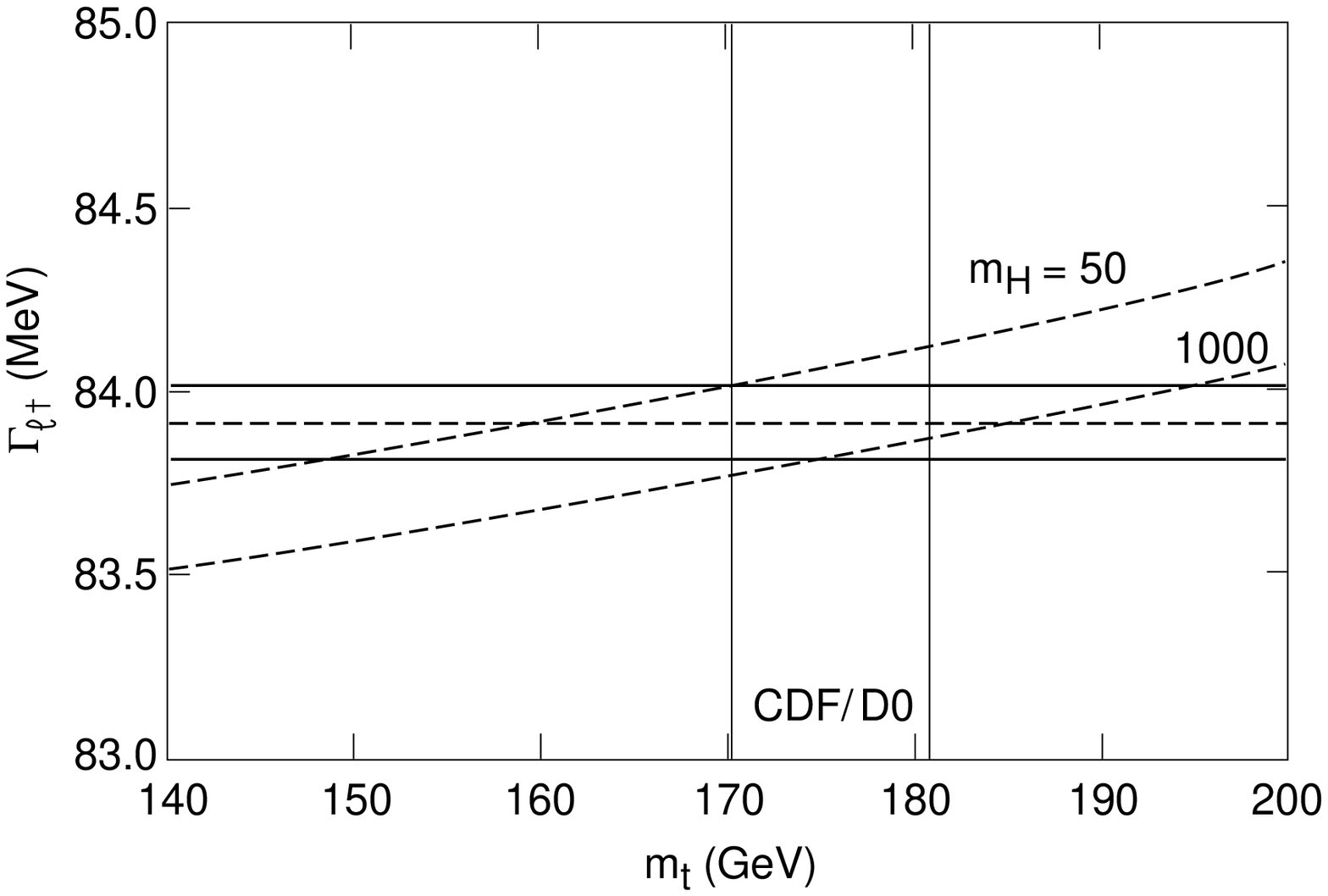}
\center{Fig. 4}
\end{figure}
\begin{figure}[t]
\psfig{figure=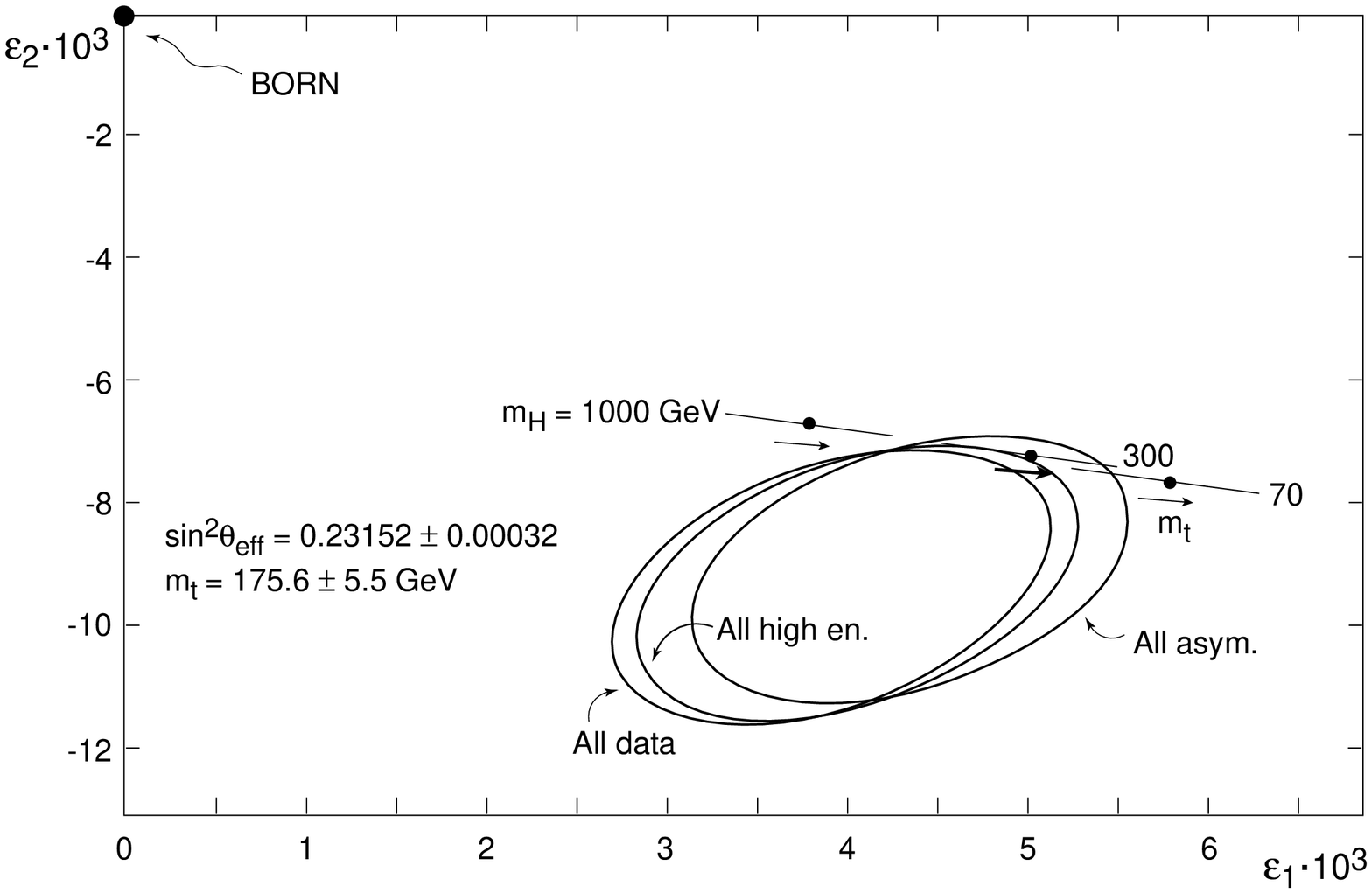,width=18cm}
\center{Fig. 5}
\end{figure}
\begin{figure}[t]
\psfig{figure=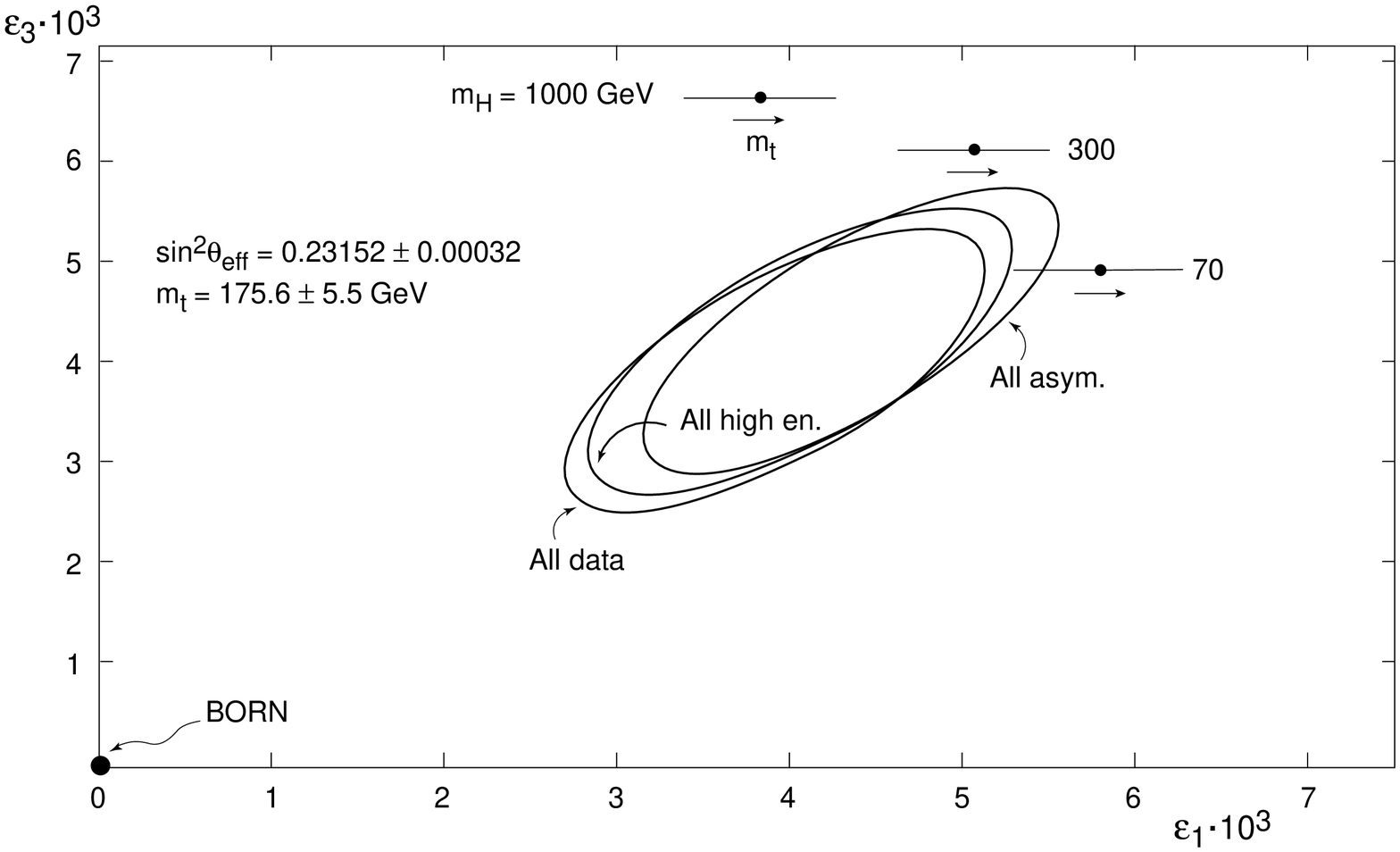,width=18cm}
\center{Fig. 6}
\end{figure}

\begin{figure}[t]
\psfig{figure=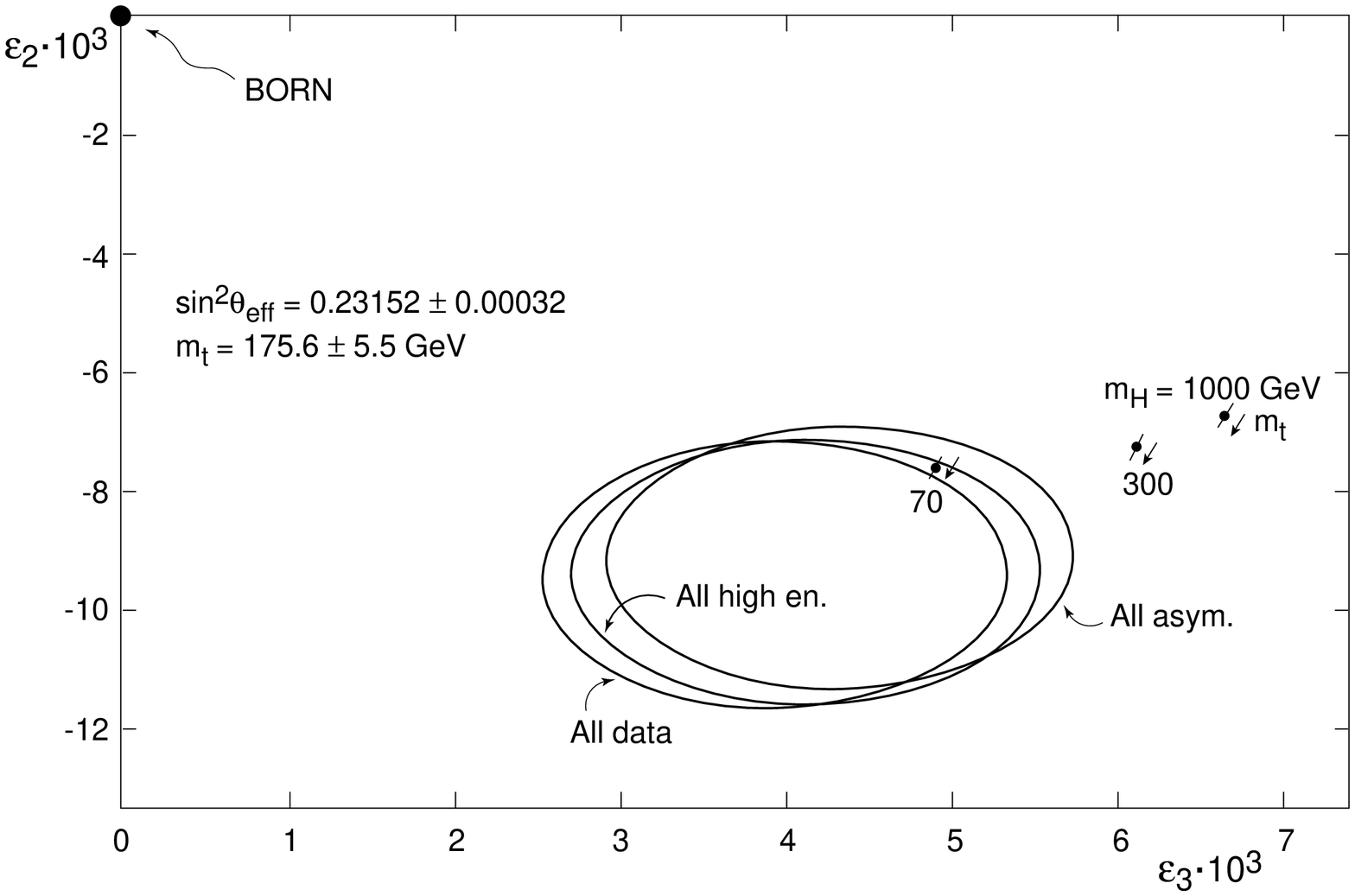,width=18cm}
\center{Fig. 7}
\end{figure}
\begin{figure}[t]
\psfig{figure=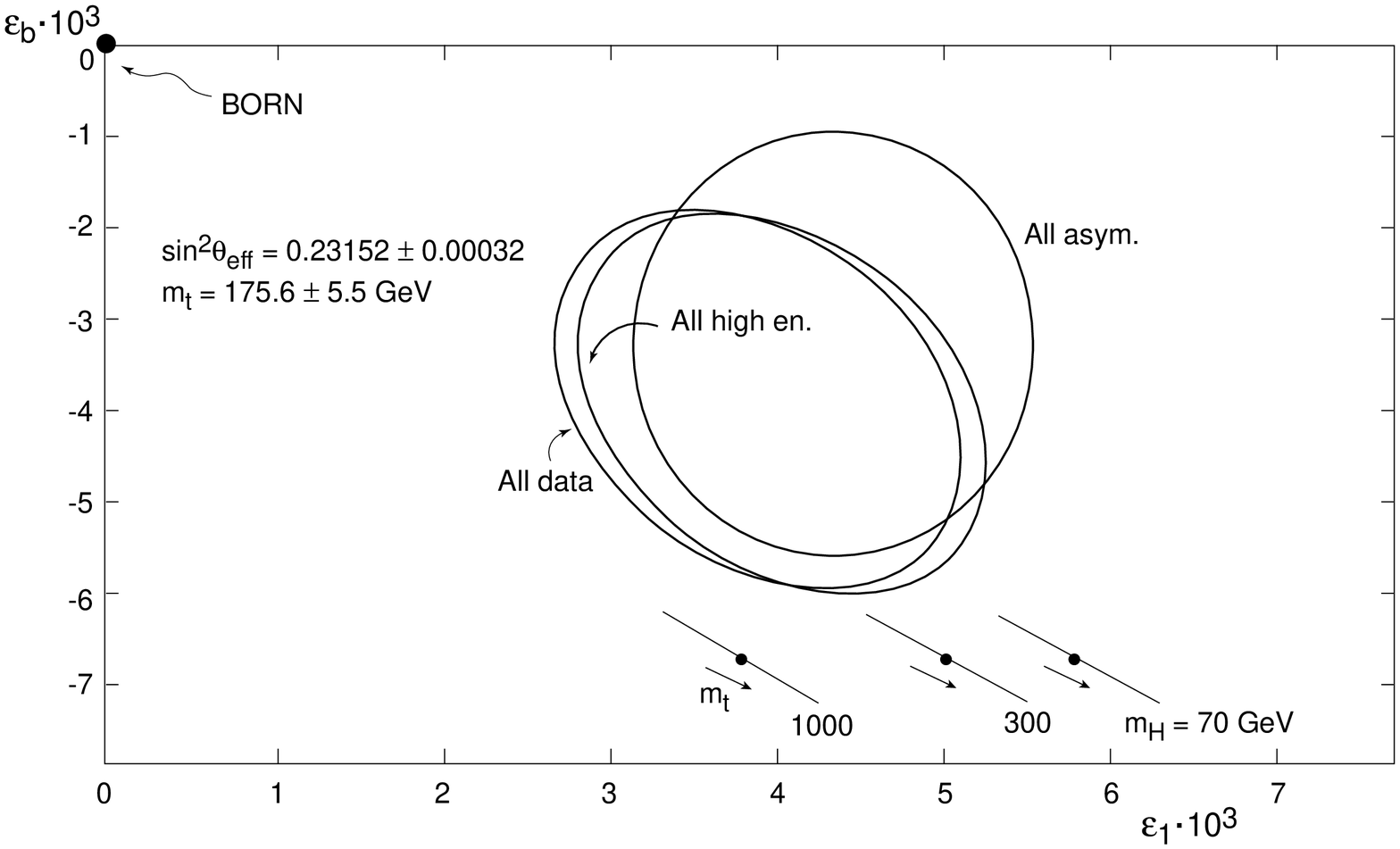,width=15cm}
\center{Fig. 8}
\end{figure}
\begin{figure}[t]
\psfig{figure=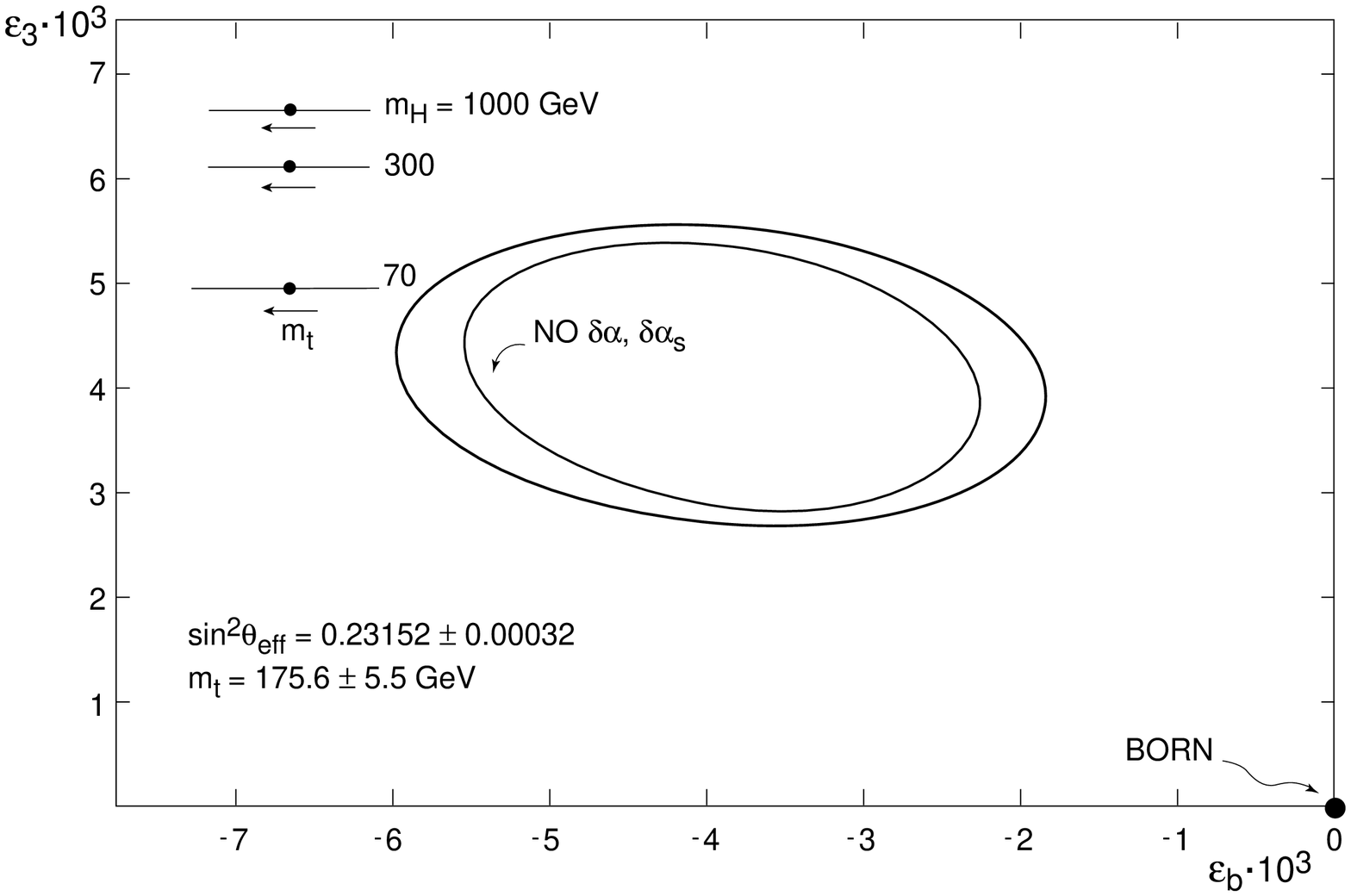,width=18cm}
\center{Fig. 9}
\end{figure}

\begin{figure}[t]
\psfig{figure=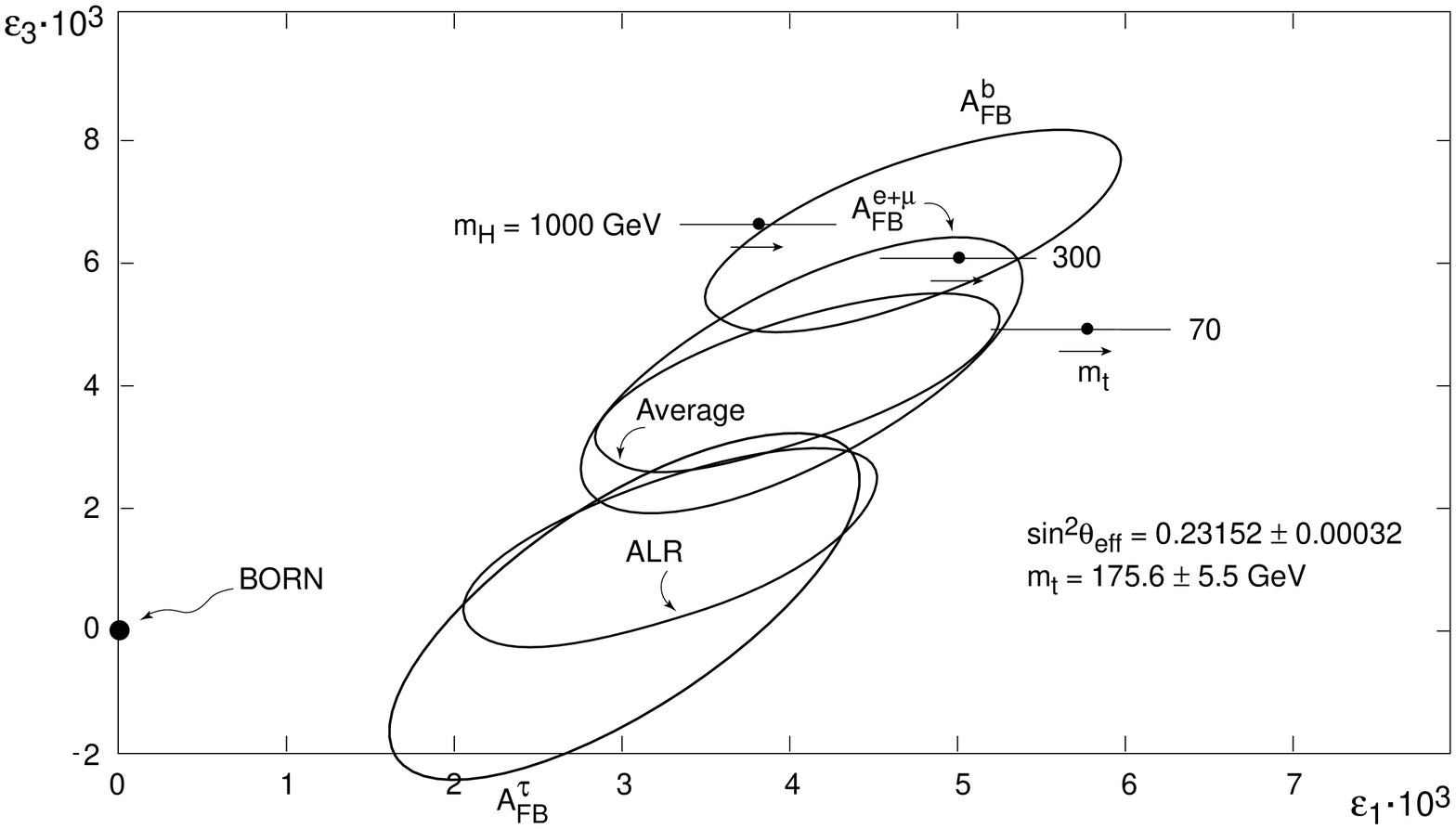,width=18cm}
\center{Fig. 10}
\end{figure}
\begin{figure}[t]
\psfig{figure=Altarelli2.eps.nbl,width=18cm}
\center{Fig. 11}
\end{figure}
\begin{figure}[t]
\psfig{figure=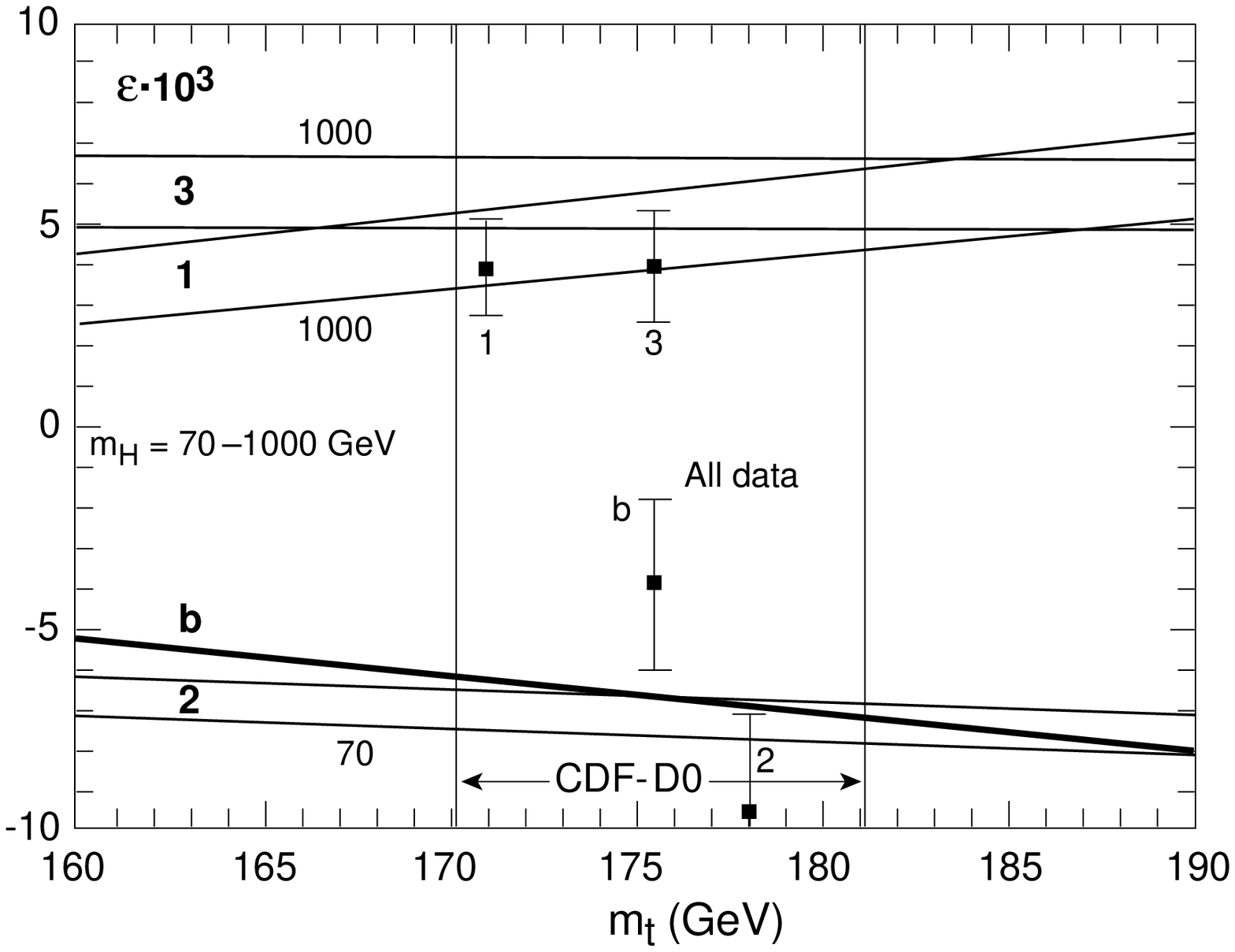,width=18cm}
\center{Fig. 12}
\end{figure}
\begin{figure}[t]
\psfig{figure=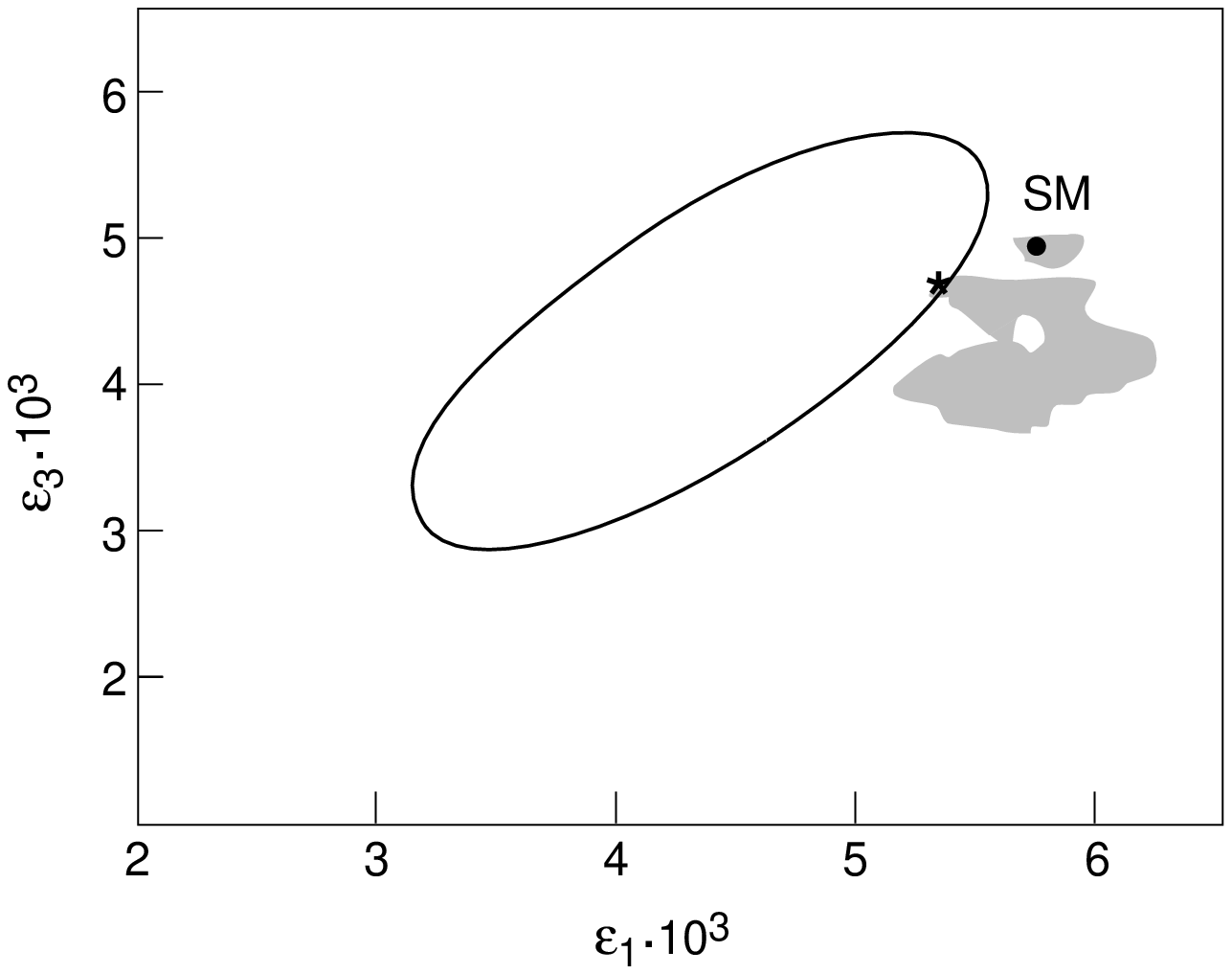,width=18cm}
\center{Fig. 13}
\end{figure}
\begin{figure}[t]
\psfig{figure=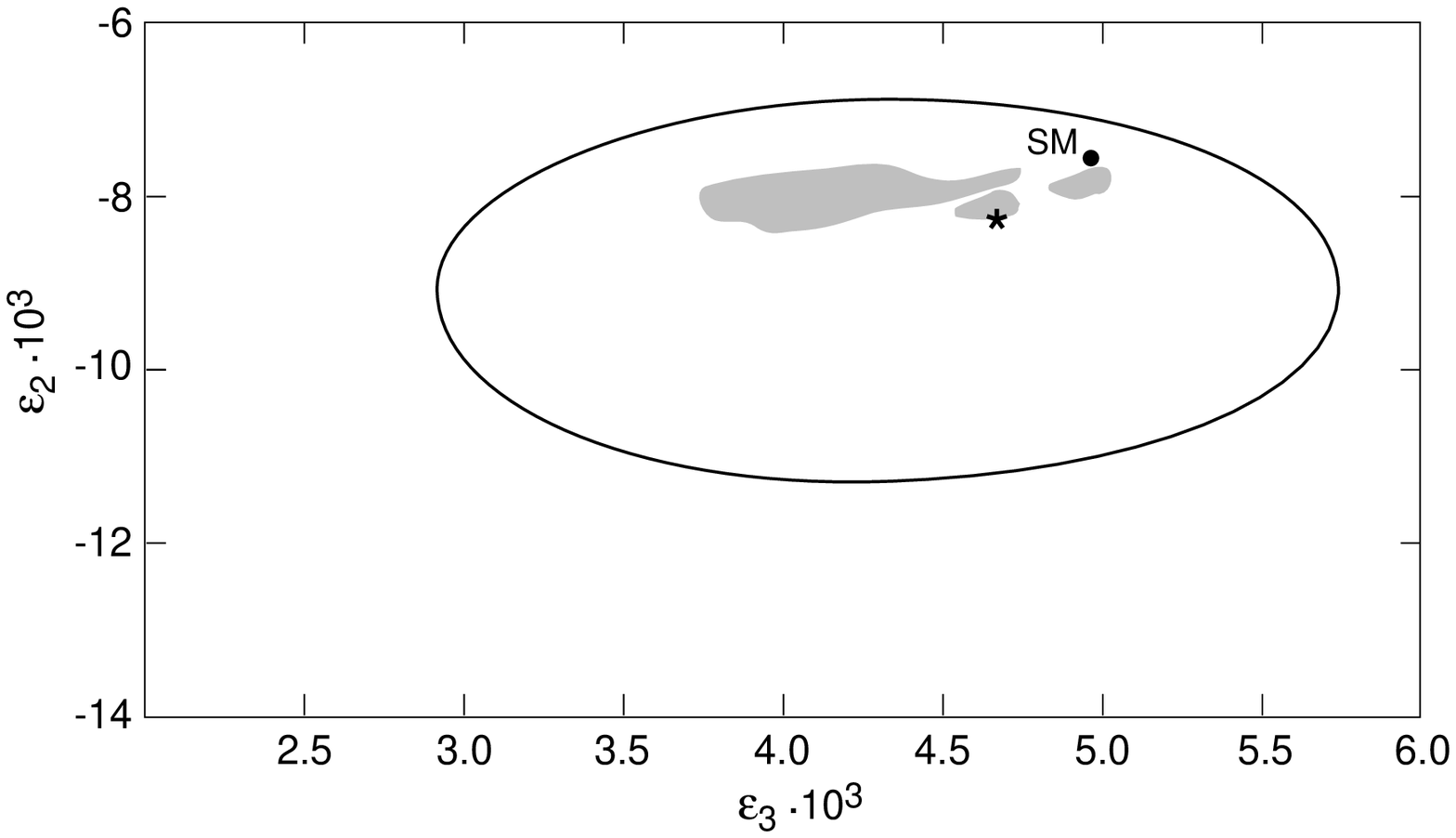,width=18cm}
\center{Fig. 14}
\end{figure}

\begin{thebibliography}{199}
\bibitem{tim} J.Timmermans, Proceedings
of LP'97, Hamburg, 1997;  S. Dong, ibidem;  D.Ward, Proceedings
of HEP97, Jerusalem, 1997.
\bibitem{ew} The LEP Electroweak Working Group, LEPEWWG/97-02.
\bibitem{dio} C.Dionisi, Proceedings
of LP'97, Hamburg, 1997;  P.Janot,Proceedings
of HEP97, Jerusalem, 1997.
\bibitem{30}    S. Weinberg, Phys. Rev. D13(1976)974 and Phys. Rev. 
D19(1979)1277; 
        L. Susskind, Phys. Rev. D20(1979)2619; 
        E. Farhi and L. Susskind, Phys. Rep. 74(1981)277.
\bibitem{31}    R. Casalbuoni et al., Phys. Lett. B258(1991)161; 
        R.N. Cahn and M. Suzuki, LBL-30351 (1991); 
        C. Roisnel and Tran N. Truong, Phys. Lett. B253(1991)439; 
        T. Appelquist and G. Triantaphyllou, Phys. Lett. B278(1992)345; 
        T. Appelquist, Proceedings of the Rencontres de la Vallée d'Aoste,
 La Thuile,        Italy, 1993; 
        R.Chivukula, hep-ph/9701322.
\bibitem{32}    J. Ellis, G.L. Fogli and E. Lisi, Phys. Lett. B343(1995)282.
\bibitem{li} W.Li, Proceedings
of LP'97, Hamburg, 1997.
\bibitem{car} F.Caravaglios, Phys.Lett.B394(1997)359.
\bibitem{gir} P.Giromini, Proceedings
of LP'97, Hamburg, 1997;  A. Yagil,Proceedings
of HEP97, Jerusalem, 1997.
\bibitem{radcorr}G. Altarelli, R. Kleiss and C. Verzegnassi (eds.), Z  Physics
at LEP 1 (CERN 89-   08, Geneva, 1989), Vols. 1--3;   Precision
Calculations for the Z Resonance, ed. by D.Bardin, W.Hollik and
G.Passarino, CERN Rep 95-03 (1995);  M.I. Vysotskii, V.A. Novikov, L.B. Okun
 and A.N. Rozanov, hep-ph/9606253.
\bibitem{alfaQED}       F.Jegerlehner, Z.Phys. C32(1986)195,B.W.Lynn, G.Penso
 and C.Verzegnassi, Phys.Rev.
D35(1987)42;  H.Burkhardt et al.,  Z.Phys. C43(1989)497; F.Jegerlehner, Progr.
 Part. Nucl. Phys. 27(1991)32; 
M.L.Swartz, Preprint SLAC-PUB-6710, 1994; M.L.Swartz,
 Phys.Rev.D53(1996)5268;  A.D.Martin and D.Zeppenfeld,
Phys.Letters B345(1995)558;  R.B. Nevzorov, A.V. Novikov, M.I. Vysotskii,
 hep-ph/9405390;  H.Burkhardt and
B.Pietrzyk, Phys.        Lett. B356(1995)398;    S.Eidelman and F.Jegerlehner,
  Z.Phys. C67(1995)585.
\bibitem{piet} B.Pietrzyk, Proceedings of the Symposium on Radiative
 Corrections, Cracow, 1996.
\bibitem{cat} S.Catani, Proceedings
of LP'97, Hamburg, 1997;  Yu.L. Dokshitser, Proceedings
of HEP97, Jerusalem, 1997.
\bibitem{radcorr2} ZFITTER: D. Bardin et al., CERN-TH. 6443/92 and refs.
therein;  TOPAZ0: G. Montagna et al., Nucl. Phys. B401(1993)3, Comp. Phys.
 Comm.         76(1993)328;  BHM: G.Burgers et
al., LEPTOP: A.V. Novikov, L.B.Okun and M.I. Vysotsky,
 Mod.Phys.Lett.A8(1993)2529;  WOH, W:Hollik :  see         ref.
\cite{radcorr}.
\bibitem{deg} G.Degrassi, P.Gambino and A.Vicini, Phys.
  Lett. B383(1996)219;  G.Degrassi, P.Gambino and A.Sirlin,
Phys.   Lett. B394(1997)188;  G.Degrassi, P.Gambino, M.Passera and
 A.Sirlin,hep-ph/9708311.
\bibitem{fits} J. Ellis, G.L. Fogli and E. Lisi, Phys. Lett. B389(1996)321; 
 G.Altarelli, hep-ph/9611239; 
A.Gurtu, Phys. Lett. B385(1996)415;  P.Langacker and
 J.Erler, hep-ph/9703428;  J.L.Rosner, hep-ph/9704331;  K.Hagiwara,
D.Haidt and S.Matsumoto, hep-ph/9706331.
\bibitem{kim} Y.Y.Kim, Proceedings
of LP'97, Hamburg, 1997.
\bibitem{abc}   G. Altarelli, R. Barbieri and S. Jadach, Nucl. Phys. 
B369(1992)3;  G. Altarelli, R. Barbieri and F. Caravaglios, Nucl. Phys. 
 B405(1993)3;  Phys.         Lett. B349(1995)145.
\bibitem{pes} M.E. Peskin and T. Takeuchi, Phys. Rev. Lett. 65(1990)964
and Phys. Rev. D46(1991)381.
\bibitem{abar}  G. Altarelli and R. Barbieri, Phys.
Lett. B253(1990)161;  B.W. Lynn, M.E. Peskin and R.G. Stuart, SLAC-PUB-3725
 (1985);   in Physics at     LEP, Yellow
Book CERN 86-02, Vol. I, p. 90;  B. Holdom and J. Terning, Phys. Lett. 
B247(1990)88;  D.C. Kennedy and P. Langacker,
Phys. Rev. Lett. 65(1990)2967.
\bibitem{zbb}
A.A. Akundov et al., Nucl. Phys. B276(1988)1; 
F. Diakonov and W. Wetzel, HD-THEP-88-4 (1988); 
W. Beenakker and H. Hollik, Z. Phys. C40(1988)569; 
B.W. Lynn and R.G. Stuart, Phys. Lett. B252(1990)676; 
J. Bernabeu, A. Pich and A. Santamaria, Phys. Lett. B200(1988)569; 
 Nucl. Phys. B363(1991)326.
\bibitem{kk} B.A.Kniehl and J.H.Kuhn, Phys. Lett. B224(1989)229;  Nucl.Phys.
 B329(1990)547.
\bibitem{carLR} F. Caravaglios and G.G.Ross, Phys.      Lett. B346(1995)159.
\bibitem{33}    CHARM Collaboration, J.V. Allaby et al., Phys. Lett. 
B177(1986)446;   Z. Phys.   C36(1987)611; 
        CDHS Collaboration, H. Abramowicz et al., Phys. Rev. Lett. 
57(1986)298; 
        A. Blondel et al., Z. Phys.  C45 (1990) 361; 
        CCFR Collaboration, K.McFarland,  hep-ex/9701010.
\bibitem{34} C.S.Wood et al., Science 275(1997)1759.
\bibitem{35}    CHARM II Collaboration, P.Vilain et al., Phys. Lett.
 B335(1997)246.
\bibitem{39} 
S. Dittmaier, D. Schildknecht, K. Kolodziej, M. Kuroda  
Nucl. Phys. B426(1994)249;  B446(1995)334; 
A. Sirlin, and P.  Gambino, Phys. Rev. Lett. 73(1994)621; 
S. Dittmaier, D. Schildknecht and G.Weiglein, Nucl. Phys. B465(1996)3.
\bibitem{43} H.P. Nilles, {    Phys. Rep.} {    C110}(1984)1; \\
H.E. Haber and G.L. Kane, {    Phys. Rep.} {    C117}(1985)75; \\
R. Barbieri, {    Riv. Nuovo Cim.} {    11}(1988)1.
\bibitem{58}    R. Barbieri, F. Caravaglios and M. Frigeni, Phys. Lett. 
B279(1992)169.
\bibitem{pok} S.Pokorski, Proceedings
of ICHEP'96, Warsaw, 1996; see also P. Chankowski, J. Ellis and S. Pokorski, 
 CERN preprint TH/97-343, hep-ph/9712234.
\bibitem{59}    R. Barbieri and L. Maiani, Nucl. Phys. B224(1983)32; \\
        L. Alvarez-Gaum\'e, J. Polchinski and M. Wise, Nucl. Phys. 
B221(1983)495.
\bibitem{60}    W. Hollik, Mod. Phys. Lett. A5(1990)1909.
\bibitem{61}    A. Djouadi et al., Nucl. Phys. B349(1991)48; \\
        M. Boulware and D. Finnell, Phys. Rev. D44(1991)2054.  The sign
 discrepancy    between these
two papers appears now to be solved in favour of the second one.
\bibitem{63}    G. Altarelli, R. Barbieri and F. Caravaglios,
 Phys. Lett. B314(1993)357.
\bibitem{zwi} A. Brignole, F. Feruglio and F. Zwirner {    Z. Phys.} 
{    C71}(1996)679.
\bibitem{zziii}         M. Sher, {    Phys. Rep.} {    179}(1989)273; 
{    Phys. Lett.}
{    B317}(1993)159.
\bibitem{aaiiii} G. Altarelli and G. Isidori, {    Phys. Lett.} {    
B337}(1994)141.
\bibitem{bbiiii} J.A. Casas, J.R. Espinosa and M. Quiros,  {    Phys. Lett.}
{    B342}(1995)171.
\bibitem{cciiii} J.A. Casas et al., {    Nucl. Phys.} {    B436}(1995)3; 
  E{    B439}(1995)466.
\bibitem{ddiiii} M. Carena and C.E.M. Wagner, {    Nucl. Phys.} {    
B452}(1995)45.
\bibitem{eeiiii} See, for example, M. Lindner,  {    Z. Phys.} {    
31}(1986)295 and references therein.
\bibitem{hr} T.Hambye and K.Riesselmann, Phys. Rev. D55(1997)7255.
\bibitem{ffiiii} H. Haber and R. Hempfling, {    Phys. Rev. Lett.} {    
66}(1991)1815;  J. Ellis, G. Ridolfi and F. Zwirner, {    Phys. Lett.} 
{    B257}(1991)83;  Y. Okado, M. Yamaguchi and T. Yanagida, {    Progr. Theor. Phys. Lett.}
{    85}(1991)1; 
        R. Barbieri, F. Caravaglios and M. Frigeni, {    Phys. Lett.}
 {    B258}(1991)167.
        For a 2-loop improvement, see also:
R. Hempfling and A.H. Hoang, {    Phys. Lett.} {    B331}(1994)99.
\end{thebibliography}
\end{document}